\documentclass[11pt,a4paper]{article}
\pdfoutput=1
\usepackage{jcappub}

\newcommand{\beq}{\begin{equation}}
\newcommand{\eeq}{\end{equation}}
\usepackage{graphicx}
\usepackage{natbib}
\usepackage{soul}

\usepackage{amsmath}

\newcommand{\be}{\begin{equation}}
\newcommand{\ee}{\end{equation}}
\newcommand{\bea}{\begin{eqnarray}}
\newcommand{\eea}{\end{eqnarray}}
\newcommand{\bi}{\begin{itemize}}
\newcommand{\ei}{\end{itemize}}

\title{Nambu-Goto Dynamics of Field Theory Cosmic String Loops}

\author[a,b,c]{Jose J. Blanco-Pillado,}
\author[a,b]{Daniel Jim\'enez-Aguilar,}
\author[a,b]{Joanes Lizarraga,}
\author[a,b,d]{Asier Lopez-Eiguren,}
\author[d]{Ken D. Olum,}
\author[a,b]{Ander Urio,}
\author[a,b]{and Jon Urrestilla}

\affiliation[a]{Department of Physics, UPV/EHU, 48080, Bilbao, Spain}
\affiliation[b]{EHU Quantum Center, University of Basque Country, UPV/EHU}
\affiliation[c]{IKERBASQUE, Basque Foundation for Science, 48011, Bilbao, Spain}
\affiliation[d]{Institute of Cosmology, Department of Physics and Astronomy, Tufts University, Medford, MA 02155, USA}

\emailAdd{josejuan.blanco@ehu.eus}
\emailAdd{daniel.jimenez@ehu.eus}
\emailAdd{joanes.lizarraga@ehu.eus}
\emailAdd{asier.lopez@ehu.eus}
\emailAdd{kdo@cosmos.phy.tufts.edu}
\emailAdd{ander.urio@ehu.eus}
\emailAdd{jon.urrestilla@ehu.eus}

\abstract{
We perform a detailed comparison of the dynamics of cosmic string
loops obtained in cosmological field theory simulations with their expected motion according to the
Nambu-Goto action. We demonstrate that these loops follow the 
trajectories predicted within the NG effective theory except in regions
of high curvature where energy is emitted from the loop in the form of massive
radiation. This energy loss continues for all the loops 
studied in this simulation until they self-intersect or become small enough that they annihilate 
and disappear well before they complete a single oscillation. We comment on the relevance of this
investigation to the interpretation of the results from cosmological field theory simulations
as well as their extrapolation to a cosmological context.
}

\begin{document}

\maketitle

\section{Introduction}

Many non-linear field theories have in their spectrum solitonic
solutions that play a significant role in the dynamics of the
theory. These objects have been extensively studied over the past few
decades in connection to many different fields of research from
condensed matter to cosmology. One important aspect of these studies
has been the investigation of the dynamics and interactions of these
solitons.

One can explore these issues by performing numerical experiments in a lattice 
simulation. This lattice does not know anything about the solitonic configuration 
and simply evolves every point in space as any other one following the equations 
of motion. One can then recover the information of the evolution of the fields in the 
lattice and interpret the results in terms of a collection of solitons 
in motion possibly interacting with each other. This is a useful way to analyse the 
microphysics of the solitons since one can probe the profiles of the fields at scales 
smaller than the characteristic soliton size\footnote{This is always the case since 
one of the requirements of our lattice spacing should be that it is always smaller than 
the soliton characteristic scale.}.  In particular, one could use this technique to find 
out whether the solitons remain in their lowest energy state or whether they are 
typically excited at some point during the simulation.

On the other hand, this procedure is computationally very expensive if
one wants to simulate a volume that is large compared to the size of
the soliton. This is particularly relevant in simulations that involve
many solitons or whenever one is interested in investigating an effect
whose typical scale is much larger than the soliton's thickness. In
these cases one is forced to look for some effective theory that
captures the degrees of freedom that are relevant for the problem
without having to simulate every point in a lattice. This drastic
reduction in the number of degrees of freedom that one needs to
compute to simulate the soliton's dynamics suggests the possibility of
another kind of simulation. Such simulations based on effective
theories allow for a much larger dynamic range in the simulation,
which in turn will help us obtain a better understanding of the
large scale dynamics of the problem. However, one needs to make sure
that there are no microphysical effects that are missing in the
effective theory that could potentially become relevant for the large
scale dynamics that one wants to faithfully reproduce in the
simulation.

These two types of simulations are therefore complementary. One can use the
lattice simulations to learn the important field theory effects that need to 
be accounted for in the effective theory of the solitons. Once this is done, one
should be able to find some common ground where both these simulations can
be compared and where an agreement can be reached on the important dynamics 
to study. Once this is achieved, an extrapolation to the interesting scales can be 
safely done using the effective theory.

In this paper we would like to take the first step towards showing
this agreement between these two approaches in the
context of local cosmic string networks. In this case, we will
consider the Abelian-Higgs model as the field theory where local
cosmic strings occur as solitons \cite{NIELSEN197345} and the
Nambu-Goto (NG) action as the effective theory at low energies
\cite{Nambu70,Goto:1971ce}. Cosmological simulations of both types,
lattice field theory
\cite{Vincent:1997cx,Moore:2001px,Bevis:2006mj,Hindmarsh:2008dw,Hindmarsh:2017qff,Helfer:2018qgv,Correia:2020yqg,Correia:2021tok}
and Nambu-Goto dynamics \cite{Albrecht:1984xv, Bennett:1989ak,
  Allen:1990tv, Vanchurin:2005pa,Vanchurin:2005yb, Martins:2005es, Ringeval:2005kr,Olum:2006ix,
  Blanco-Pillado:2011egf,Blanco-Pillado:2013qja}, have been extensively studied in the
past. However, there seems to be an important disagreement about the
abundance of non-self-intersecting (NSI) loops between these two numerical
techniques. As its name suggests, non-self-intersecting loops are loops that in their dynamics do not self-intersect. All the loops found so far in field theory simulations do self intersect, and the loops seem to decay in a short period of time. Nambu-Goto simulations also produce self-intersecting loops, but over the course of their
evolution a large number of non-self-intersecting loops appear. These non-self-intersecting loops are important, since their main energy loss mechanism is via gravitational waves, whereas loops that continually intersect lose energy also via massive radiation.  

This apparent disagreement
has been the source of a debate for a very long time, since some of
the first field theory simulations of string networks
\cite{Vincent:1997cx}. One possible explanation could be that field theory simulations have not been lucky enough to produce a large enough NSI loop. This is a reasonable explanation, since field theory simulations have much fewer loops in general, and Nambu-Goto dynamics shows that NSI conditions in a random loop are rare compared to self-intersecting ones.

It is important to remember at this point that the NG action is only an approximation 
to the actual dynamics of the strings and, as we will describe in detail in this
paper, can certainly break down under some special circumstances.
However, field theory simulations of individual smooth strings have shown conclusively
that these strings follow almost exactly the Nambu-Goto dynamics \cite{Olum:1998ag,Olum:1999sg}.
This makes this disagreement with field theory simulations more puzzling and has prompted some authors to suggest that another reason for
the disagreement may be due to the presence of excitations on the strings in the network simulations
\cite{Hindmarsh:2017qff,Hindmarsh:2021mnl}. 

These field theory excitations have been found in many soliton solutions,
in particular in cosmic strings \cite{Arodz:1991ws,Goodband:1995rt,Alonso-Izquierdo:2015tta,Kojo:2007bk},
and have been recently studied in a series of papers in different models in 
\cite{Blanco-Pillado:2020smt,Blanco-Pillado:2022rad,Blanco-Pillado:2022axf}. The results 
of these studies show that many solitonic solutions may store a significant amount of
energy in the form of excitations \cite{Blanco-Pillado:2020smt,Blanco-Pillado:2021jad}. In particular, the phase
transition that creates these solitons could lead to an excited initial
state due to the extra energy floating around the bulk as the soliton
is formed. Furthermore, these papers show that the time scale for the
decay of these excitations could be much larger than the typical time
scale of the soliton, the light-crossing time of the width of the soliton \cite{Manton:1996ex,Blanco-Pillado:2020smt}. This
suggests that the effect of these excitations could play some role
in the dynamics of strings in field theory simulations. In particular,
the presence of these extra modes localized on the string could
modify their equation of state rendering their dynamics quite
different from the one predicted by the Nambu-Goto action. However, the existence
of this extra energy on cosmic strings has never been conclusively shown 
in any field theory simulation of the string network\footnote{The closest study of this kind
in the literature is the one presented in \cite{Blanco-Pillado:2021jad} where the authors perform
cosmological simulations of $2+1$ dimensional global vortices. Their results indicate 
that the amount of energy in this case is rather low.}.

In this paper we will study this problem 
by carefully analysing the evolution of loops extracted directly from 
field theory simulations. Specifically, we obtain the position and
velocity of some of the largest loops found in the course of a field theory
simulation of a network of strings and compare their evolution with
the one predicted by the Nambu-Goto action. 

The results indicate that these loops follow the same trajectories as their
NG counterparts, except in localized regions where the curvature of
the strings is large compared with the string core thickness, where the NG approximation is not good by definition. As we will
argue in the main part of the text, this shows that there is no significant 
deviation from the Nambu-Goto dynamics due to a new equation of state
for the strings in the parts where the curvature is not high. In other words, it seems that the strings in our simulations do not
have a large amount of energy stored in them in the form of localized 
excitations.

The organization of the paper is the following. In
section~\ref{sec:FT}, we comment on the characteristics of the field
theory simulations that were used to generate the field theory
trajectories of the loops. In section~\ref{sec:comparison}, we
describe the techniques we use to compare the dynamics of field theory
loops and their Nambu-Goto predictions. In section~\ref{sec:results},
we show our results with a few snapshots of the string trajectories
comparing both field theory and Nambu-Goto. Finally, in
section~\ref{sec:conclusion}, we comment on the implications of these
results for the cosmological extrapolation of field theory cosmic
string networks.

\section{Field theory simulations of cosmic string loops}
\label{sec:FT}
The field theory that we will investigate in this paper is the Abelian-Higgs
model, whose Lagrangian density,
\beq
L = D_{\mu} \phi D^{\mu} \phi^* - \frac{\lambda}{4} \left(|\phi|^2 - \eta^2\right)^2- \frac{1}{4e^2} F^{\mu \nu} F_{\mu \nu}~,
\label{lag}
\eeq
describes the dynamics of a complex scalar field, $\phi(x)$, coupled
to a vector field, $A_{\mu}(x)$, through the covariant derivative, 
$D_{\mu} \phi = (\partial_{\mu} - i A_{\mu}) \phi $. Furthermore, the
usual field strength for the vector field is given by $F_{\mu \nu} = \partial_{\mu} A_{\nu} -  \partial_{\nu} A_{\mu}$.
We will consider the case $\beta = \lambda/ (2e^2) = 1$, which 
means that the masses of the excitations in the vacuum for the vector 
and scalar fields are equal: $m = m_s = m_v$.

It is well known that the equations of motion obtained from this theory
allow for the existence of solitonic field theory vortices  \cite{NIELSEN197345}. In $3+1$ 
dimensions the vortices become strings whose energy is concentrated
in a core thickness of the order of $\delta \sim m^{-1}$.

In order to investigate the loops in this theory we follow the
prescription detailed in \cite{Hindmarsh:2021mnl}, where network loops
are created from some random initial conditions in the lattice. We point the interested reader to  \cite{Hindmarsh:2021mnl} for details, but we summarize here the necessary basic information.  After discretization of the Hamiltonian that corresponds to the Lagrangian (\ref{lag}), we obtain the equations of motion and solve them  in cubic lattices with periodic boundary conditions.

The initial configuration of the system is chosen to be such that all fields are set to zero except for the scalar field, which is set to be a stationary Gaussian random field with a power spectrum given by
\begin{equation}
	P_{\phi}=Ae^{-k l_{\phi}} \; ,
\end{equation}
where the amplitude $A$ is chosen so that $\left\langle \left |\phi \right |^{2} \right \rangle = \eta^{2}$. The free parameter which rules the initial randomness is the correlation length $l_{\phi}$, which can be set initially to different values.

These random initial conditions lead to a considerable excess of energy in the simulation volume. Therefore, a cooling process is applied using a diffusive period of evolution, and once a smooth field distribution is obtained, the network evolves following the true equations of motion in flat space. 

It is important to mention that after the diffusive period  the network is at rest, which means that any loop at this initial stage will start from a static configuration. We will comment on these primordial loops in Appendix \ref{ploops}, but let us for now comment on the fact that they are not the main interest to us since they are not representative of the typical loops in a network simulation. In the course of the evolution of the network, loops of strings will be formed by either self-intersections or intercommutation of long strings and these are the  loops that we are mainly interested in. By this stage the string is in motion and so the loops formed are not static. Note that the numerical evolution of the string network was performed in flat spacetime, in other words on a Minkowski lattice spacetime.

In order to localize the position of the strings and the subsequent loops we first identify all the plaquettes with a non-trivial winding in our simulations. In this way the connection between all the centers of these plaquettes will constitute the string. We follow the evolution of the network outputting  the windings and confirm the formation of loops coming from intersections  by visual inspection. 

Since, as mentioned, the simulations are done using periodic boundary conditions, all the strings in the box can be considered to be closed loops; but these loops can be ``broken'' by the periodic boundary conditions. Thus, if one were to plot the loop directly, some would not appear to be a connected piece of string.  In order to avoid this we reconstruct the loops that are broken by the periodic boundary conditions applying spatial translations and assigning new coordinates to the string positions. In this manner, we have a list of connected positions in space for all loops.

In total we have analyzed 7 loops (and their descendants) from the simulations in  \cite{Hindmarsh:2021mnl}. These loops were obtained using two different correlation lengths, $l_\phi=15$ and $25$ in $\eta^{-1}$ units. All of them were produced using lattices of $N=1024$ points per dimension with a spatial resolution of $\delta x=0.125$ and temporal resolution of $\delta t = 0.2\delta x$, again in $\eta^{-1}$ units. We refer the reader to \cite{Hindmarsh:2021mnl} for more specific details on the preparation of these simulations. Moreover, the loops, which are reconstructed following the above prescription, are output at each time step of the evolution so that the field theory information available for the Nambu-Goto prediction is  the most accurate possible. 
Furthermore, as explained in \cite{Hindmarsh:2021mnl} by the time the loops we use here get formed, the large scale dynamics of the string network is consistent with a scaling regime \footnote{Recall, however, that the simulation here is done in flat space. It would be very interesting to analyze loops from a lattice simulation on an expanding background. Some effort in this direction is already underway.}.

This data, extracted directly from the field theory simulations, is the starting point of our analysis.

\section{Comparison of Field Theory data with the Nambu-Goto action}
\label{sec:comparison}

As we said above, the effective action that describes the dynamics of
local strings is expected to be the Nambu-Goto action
\cite{Nambu70,Goto:1971ce}.  This can be justified by making a judicious
choice of coordinate system around the center of the string and
integrating the action along the transverse directions of the string
\cite{Forster:1974ga}: the resulting action can be shown to be of the
Nambu-Goto form. This argument rests on several assumptions that we
now list in detail.

First, it assumes that the local curvature of the string is small
compared to its thickness.  In fact, one can consider the Nambu-Goto
action as the lowest order approximation of an infinite expansion in
terms of the ratio of $\delta/R$, where $\delta$ is the nominal
thickness of the string soliton and $R$ is the radius of curvature of
the string in space. In a cosmological setting, truncating this series
keeping only the first term seems quite reasonable since the
separation of scales from the microphysical size of the strings to any
cosmologically relevant scale is phenomenally large. Of course, this
separation of scales is not so large in a field theory simulation of a
string network.

The second assumption, which is somewhat related to the previous one, is
that the string does not lose energy by radiation in the course of its
evolution.  This is of course built in the NG action since, as we
discuss below, there is a conservation law for the invariant energy of
a loop. However, from the point of view of field theory, one might
imagine a situation where solitonic strings lose part of their energy
into radiation in the form of propagating modes in the bulk.  Of
course this cannot happen for a relaxed static string, since by
definition this object is the lowest energy configuration with the
particular boundary conditions, the winding of the scalar
field. Boosting this object can not lead to radiation either. So the
only way this string can radiate is due to acceleration. This is easy
to achieve in strings since during their evolution they can develop
regions of curvature that will induce acceleration. The question is
then a quantitative one. How much energy is radiated from the typical
acceleration present in the evolution of strings? In order to answer
this question we should remind ourselves that all the propagating
modes in this model of local strings are massive. This suggests that
one should wiggle the string with a frequency at least of the order of
this mass, $m$ in our case, if one wants to produce any
radiation. Below this frequency, the source for the radiation does not
have a large enough frequency to produce propagating modes. This
argument has been extensively used in the past and demonstrated
explicitly in numerical simulations of strings in \cite{Olum:1999sg}
and more recently in the analogous situation in domain wall strings in
\cite{Blanco-Pillado:2022rad}\footnote{In fact, the situation is a
  little more complicated since the thickness of the source (the
  oscillating string) is typically also of the order of the inverse of
  the mass of the radiated particle, namely, $\delta \sim
  m^{-1}$. This means that for higher frequencies the radiation is
  also cut-off due to interference effects
  \cite{Blanco-Pillado:2022rad}.}.

There are of course moments where the string can release part of its
energy. The simplest way to visualize this is in the lower dimensional
process of vortex-antivortex annihilation. In these events, it is
clear that the arguments leading to the conclusions above do not apply
since the topological stability of the solitons disappears. Extending
this to $3+1$ dimensions, we can classify other instances where
similar processes occur. A clear example of these kinds of events is
the interaction between two long strings, the so-called
intercommutation process by which string loops can detach themselves
from long strings \cite{Matzner:1988qqj}. Of course, this process
cannot be described by the NG action. Other processes closely related
to these are cusp formation and kink-kink collisions. During the
formation of the cusp, part of the string annihilates with itself
releasing energy in the process \cite{Olum:1998ag}. Similarly,
kink-kink collisions \cite{Matsunami:2019fss} or in general the
appearance of very high curvature regions
\cite{Olum:1999sg,Blanco-Pillado:2022rad} lead to a similar energy
ejection from the string. All these non-perturbative processes are by now well understood and need to be accounted for separately from the
NG evolution.

Finally, another important assumption in the use of the NG action to
describe strings is the general expectation that, in their rest frame,
the solitonic strings would be well approximated by the static
solution of lowest energy. The underlying idea for this expectation is
the supposition that excitations on the string will decay in a time
scale of the order of the light-crossing time of the thickness of the
string. This is a very small time scale to have any relevance for the
evolution of a loop even in a field theory simulation. However, this
assumption has been recently brought into question
\cite{Hindmarsh:2021mnl} due to the existence of localized excitations
of the field theory string that can have a long lifetime
\cite{Manton:1996ex,Blanco-Pillado:2020smt,Blanco-Pillado:2021jad}. The
presence of this extra energy can change the equation of state of the
strings and so modify the trajectory of strings. These ideas
have been recently explored in several papers in lower dimensional
models with solitons
\cite{Blanco-Pillado:2020smt,Blanco-Pillado:2021jad,Blanco-Pillado:2022rad}. The
results of these studies seem to indicate that even though some of
these solitons could have some extra energy at the moment of
formation, it is difficult to see how they can achieve the necessary
significance to alter the evolution of the solitons.

Here we set out to investigate the relevance of all these possible
effects in the evolution of cosmic string loops from cosmological
network simulations. In particular, we will compare the evolution of
the string extracted from field theory following the procedure we
indicated earlier with the one that one would infer from NG
dynamics. In the following section we will explain how to obtain the
prediction of the NG dynamics from the field theory data at any moment
in time.

\subsection{The Nambu-Goto dynamics for a cosmic string loop}

The Nambu-Goto action for a relativistic string is given by
\begin{equation}
S_{NG} = -\mu \int{d^2\xi \sqrt{-\gamma}}~,
\end{equation}
where $\mu$ is the energy per unit length of the string 
and $\gamma$ is the induced metric on the worldsheet parametrized by the coordinates $\xi^{1,2}$.
The equations of motion for a string propagating in flat spacetime can be obtained from 
this action and are given by\footnote{See, for example, \cite{Vilenkin:2000jqa} for
an account of all the details of the NG action.}
\begin{equation}
{\bf \ddot x} - {\bf x''}=0 ~,
\end{equation}
where ${\bf x}(t,\sigma)$ parametrizes the position of the string and dotted and primed quantities denote their differentiation
with respect to the two worldsheet parameters, $(t,\sigma)$. Moreover,  we have also imposed the gauge conditions
\begin{equation}
    {\bf \dot x} \cdot {\bf x'}=0~,
\end{equation}
so the only physical velocity is perpendicular to the string, and
\begin{equation}
{\bf \dot x}^2 + {\bf x'}^2=1~,
\end{equation}
which means that the spacelike parameter, $\sigma$, is proportional to
the energy per unit length along the string.  These equations can be
integrated, so the most general solution is of the form
\begin{equation}
\label{NG-evolution}
{\bf x}(t,\sigma) = \frac{1}{2}\left( {\bf a} (\sigma-t) +{\bf b} (\sigma + t)  \right)~,
\end{equation}
where the constraints impose the conditions
\bea
|{\bf a'}| &=& 1 ~,\\
|{\bf b'}| &=& 1~.
\eea
This means that all one needs to do in practice to obtain the evolution
of a string is to find the form of the two functions ${\bf a} (\sigma_{-})$ and  ${\bf b} (\sigma_{+})$
with respect to their argument. In the following we will describe an algorithm to obtain these functions
from the position of the string at two different time steps. This will allow us to apply this 
procedure to the data obtained from the field theory simulation. 

Note that after obtaining these functions at a particular moment, eq.~\eqref{NG-evolution} will 
allow us to find the position of the string at any subsequent time. This can in turn be compared
with the position in field theory. If the string
moves exactly as the NG predicts, we could obtain the form of these functions
at any moment in time and the result would be the same. In the following we will
explain that it will be convenient to repeat this procedure at several times.

Another important point about the NG dynamics is the fact that one can
find the integral of the parameter $\sigma$ along the string. This is
a constant of motion of any loop of string and so can also be used
as a measure of the NG dynamics.
 
\subsection{Obtaining the NG dynamics from field theory data}

As we mentioned earlier, we extract the information about individual loops from the lattice field
theory simulation of the network by identifying their position at any moment in time. Here we 
will explain in detail how to transform that into a prediction of the NG evolution.

We start with the position of the string through the lattice as a list
of spatial positions of plaquette centers, ${\bf p}_n$,
$n=1,2,\ldots$. The first thing we do is to smooth out these position
vectors since otherwise we will have big jumps related to the discrete
nature of the lattice. One reasonable possibility is to smooth this data by
a Gaussian window function whose width is given by a few lattice
spacings, $ \text{M} \delta x$.  In order to justify this choice, let
us first remember that in order to have a faithful simulation of the
relevant dynamics the thickness of the string is somewhat larger than
the lattice spacing. Therefore, one should not expect to know the
position of the center of the strings with a precision much larger
than this width\footnote{We use $\text{M}=2$ for the loops presented in this paper. We have checked
that our results do not change significantly using $\text{M}=5$ .}.
  
Using this smoothed data, we obtain the list of vectors that describe
the normalized tangent vector of the string by computing
\beq
{\bf \hat p'_n}= \frac{{\bf p}_{n+1} - {\bf p}_{n}} {|{\bf p}_{n+1} - {\bf p}_{n}|}~.
\eeq
Now we need to compute the velocity vectors for each point of the
string. In order to do that, we will assume the NG evolution and consider
that the string moves in the direction perpendicular to its tangent vector.
This allows us to compute this velocity using the following algorithm.

First, we find the point of the string at a later time described by
$t+\Delta T$ that is the intersection of the plane perpendicular to
the tangent vector at the original position of the string (at time
$t$) with the string at $t+\Delta T$. Let's denote this point by
$\tilde {\bf p}_n(\Delta T)$.  Then we estimate the velocity vector of
this segment of the string in the NG approximation by taking
\beq
{\bf \dot x}_{n} =  (\tilde {\bf p}_n(\Delta T) - {\bf p}_{n})/\Delta T ~.\\
\eeq

Here we should comment on an important point. In the previous
algorithm we have not specified what is the relation between the time
interval between the two sets of strings positions we have in the
original data and the computing time interval in the numerical field
theory simulation, $\delta t$. In our case, we have experimented with
different values and finally settled on $\Delta T = 8~\delta t$.
The reason to take these two snapshots of the string separated in time
larger than the minimal possible time separation, $\delta t$, is to
try to smooth out possible errors in the estimate of the
velocity\footnote{Using a smaller value of $\Delta T$ induces the
  presence of points of super-luminal motion.  This is of course an
  error induced in regions of high velocity of the string. Note also
  that regions of self-annihilation would give rise to these
  problems. However, it is clear that this is to be expected since in
  those regions the string does not behave as NG, so this algorithm
  should definitely fail.}.

Using this velocity and the normalized tangent vectors we can now compute the 
tangent vector correctly parametrized according to the usual NG gauge. This means 
that we can define the new tangent vectors as 
\beq
{\bf x}'_n =  \left(\sqrt{1-|{\bf \dot x}_n|^2}\right) {\bf \hat p'_n}~.
\eeq

Using the velocity and the tangent vectors we can in turn compute the functions 
${\bf a}'$ and ${\bf b}'$ using
\begin{eqnarray}
{\bf a}'_n =  {\bf x}'_n - {\bf \dot x}_{n}\,, \\
{\bf b}'_n =  {\bf x}'_n + {\bf \dot x}_{n} \,,
\end{eqnarray}
and from here it is easy to find the position of the string at any moment
in time following the prescription of the NG solution in eq.~\eqref{NG-evolution}.

Finally, using this data we can easily compute the local Lorentz factor 
associated with each segment, namely,
\beq
\Gamma_n = \frac{1}{\sqrt{1- |{\bf \dot x}_n|^2}}~,
\eeq
as well as the amount of $\sigma$ parameter in each
segment of the string by computing
\beq
\Delta \sigma_n = \Gamma_n |{\bf x}_{n+1} - {\bf x}_{n} |~.
\eeq
Integrating this over our list of elements of the string we obtain the 
invariant energy of the loop, $E_{NG} = \mu \sum_{n}{ \Delta \sigma_n}$.

As we mentioned earlier, in the course of the reconstruction there are
points that lead to an estimation of the velocity from $|{\bf \dot
  x}_n|$ very close to the speed of light or even above it. In order
to suppress the pathological behaviour that these points could have on
the total energy we have decided to put an artificial cap to the
velocity of each individual point. We replace any estimate of
$|{\bf \dot x}_n|>0.9$ by $v_{\text{max}} =0.9$. We have tried other
regularization procedures and checked that the total energy of the loop
is not significantly affected by the different procedures.

\section{Results}
\label{sec:results}

Using the techniques we have outlined above, we can compare the
evolution of field theory loops from our simulations with the motion
predicted from the NG dynamics. Using the data from two steps
in the field theory simulation separated by $\Delta T$ since the
formation of the loop,  we build the ${\bf a}(\sigma)$ and ${\bf
  b}(\sigma)$ functions. This allows us to plot the predicted string
NG position for all times. This comparison shows that the field theory
string follows the NG solution very accurately for the large majority
of the string length. This seems to suggest that these loops extracted
directly from the simulation are not endowed with a significant amount
of extra energy as conjectured in \cite{Hindmarsh:2021mnl}, or at least, not enough to change the trajectory of the string perceptibly. A large
amount of energy in bound states would change the equation of state of
the string and its local velocity would be changed with respect to the
one obtained in NG. This is not observed for most parts of the string
length.

There are, however, regions where we see a departure of the field
theory string position when compared to NG. Many of the places where we see this departure are regions 
where the NG dynamics predicts the existence of a high curvature section on the string. As we
mentioned earlier it is therefore not a surprise that the field theory string does
not follow the NG prediction in those regions.

\begin{figure}
\begin{center}
\includegraphics[width=10cm, height=5cm]{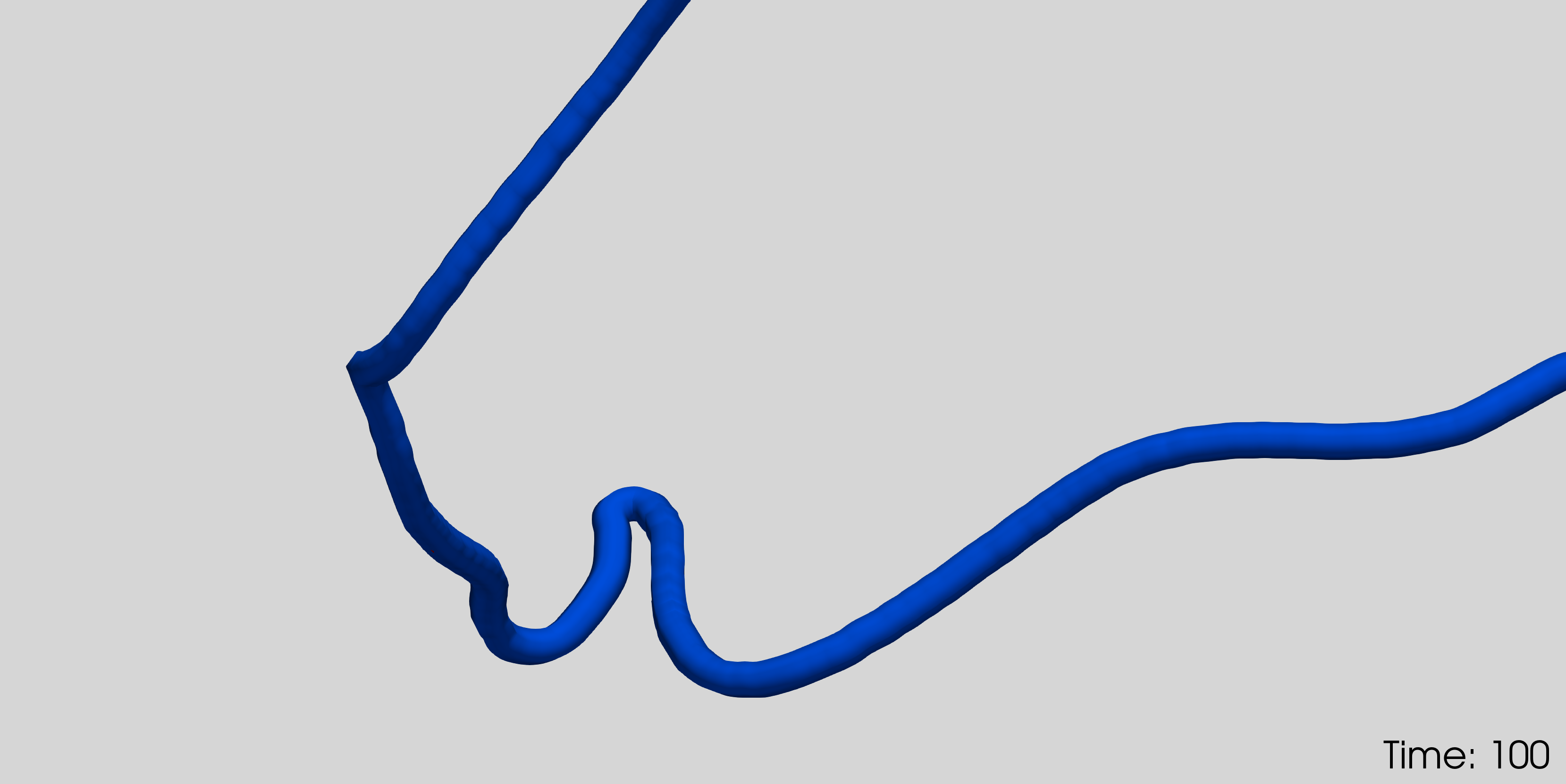}\vspace{2pt}\\
\includegraphics[width=10cm, height=5cm]{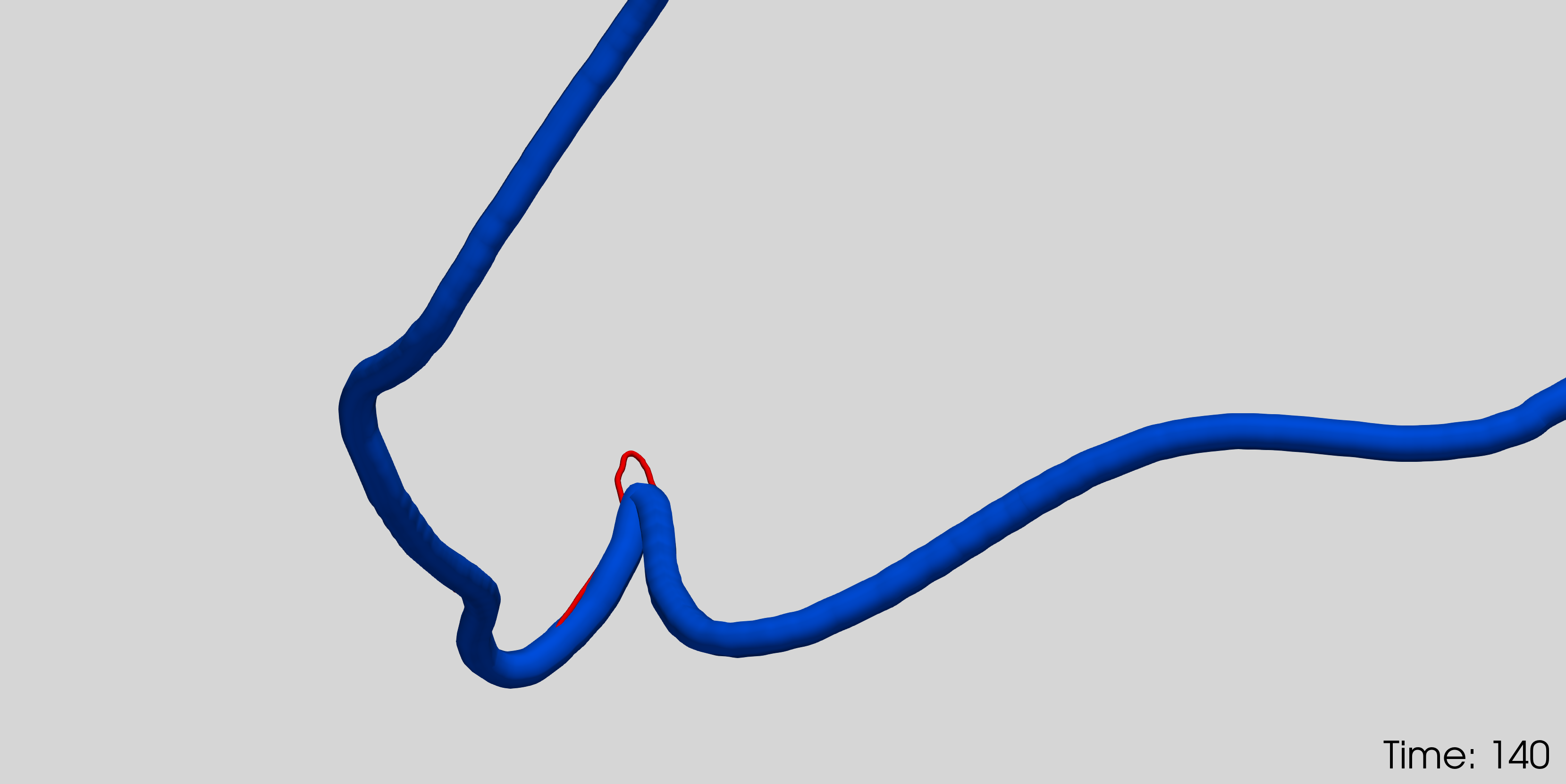}\vspace{2pt}\\
\includegraphics[width=10cm, height=5cm]{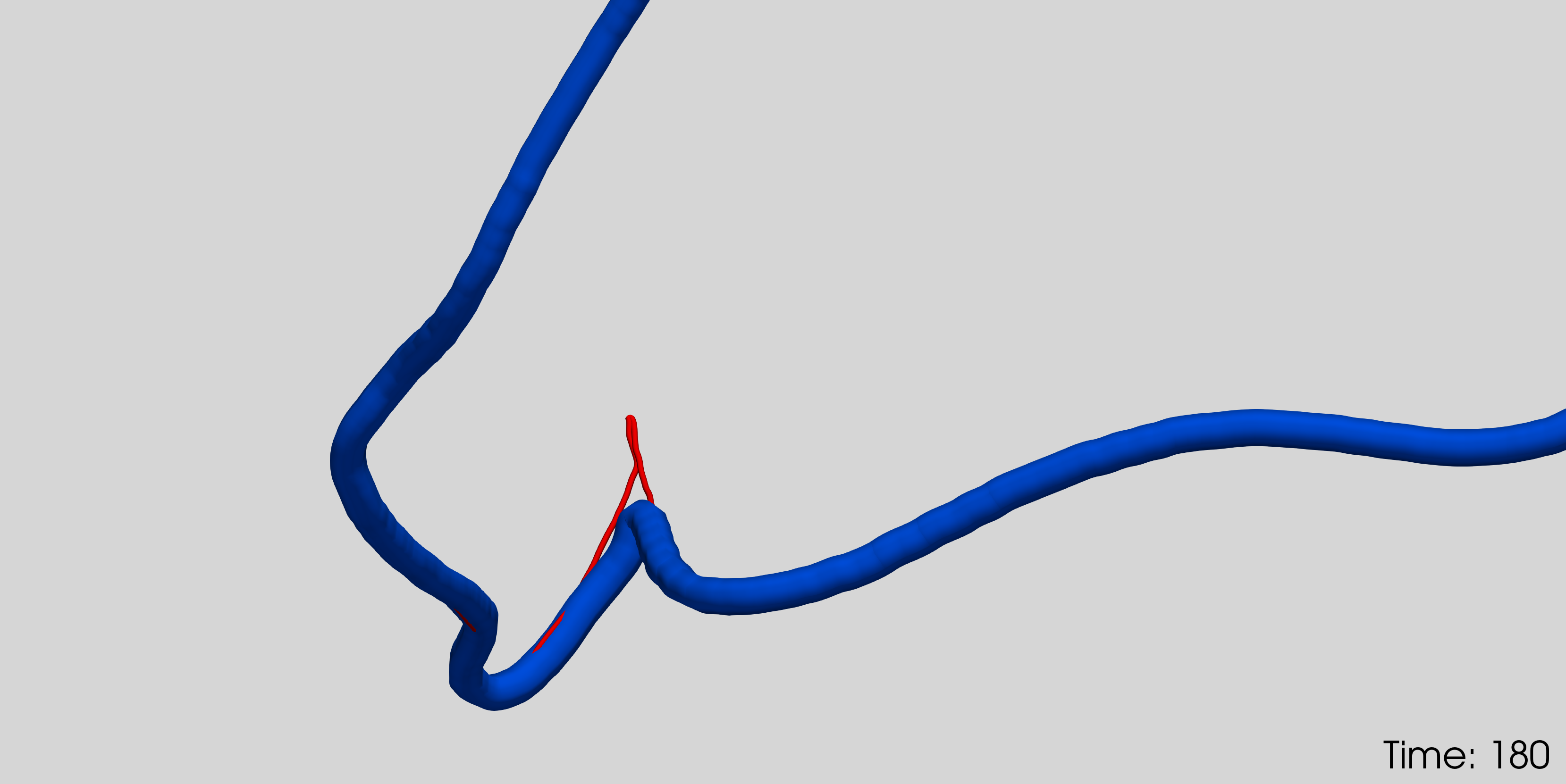}
\caption{Several snapshots of the evolution of one of the field theory
  loops. We show in blue the position of the string obtained directly
  from the lattice simulation. In red is the predicted position
  obtained from the reconstruction of the NG data at the initial
  conditions and evolved using the NG dynamics until the time
  shown. The agreement between these two descriptions is very good for
  most of the loop's evolution. We have zoomed in on a region of the
  string at a particular moment where there is a visible departure
  between them.}
\label{fig:first-event}
\end{center}
\end{figure}

An example of such local departure from NG dynamics is shown in
figure~\ref{fig:first-event}.
We present several snapshots of this loop's evolution in field theory
(in blue) and in the NG dynamics (in red) obtained using the algorithm
described in the previous section. In order to represent the field
theory string, we give it a width of the order of $\delta$ (the
thickness of the solitonic object).  The predicted position of the NG
action is hidden inside of the blue tube describing the position of
the field theory string for most of the string. The two curves only
deviate from one another in a small section of the whole string. In
that region, the NG string curves itself at the scale of the order of
the string thickness, but the field theory string does not do that and
finds a shortcut.

These episodes of high curvature act as a source of energy loss
from the string. Some of these events resemble the cusp
annihilation simulated several years ago in \cite{Olum:1998ag}. Others
just correspond to the interaction of wiggles on the string
that produce high curvature regions. In some cases, these interactions
lead to the formation of tiny daughter loops that immediately annihilate in the
field theory side. 

The subsequent evolution of the field theory string does not follow the 
NG prediction after those episodes. The reason for this is also clear. The NG action
conserves energy, and therefore it does not account for this energy
loss mechanism. This means that the evolution from field
theory would start being different in the region where energy
is radiated. As time passes, this departure from FT and NG
spreads over the rest of the string. If one waits long enough,
the difference becomes quite visible, and if one were to continue the comparison 
forward, the shapes of the loops would grow more different. However, this is not 
a real measure of  the different local dynamics. 

\begin{figure}
\begin{center}
\includegraphics[width=11cm, height=5cm]{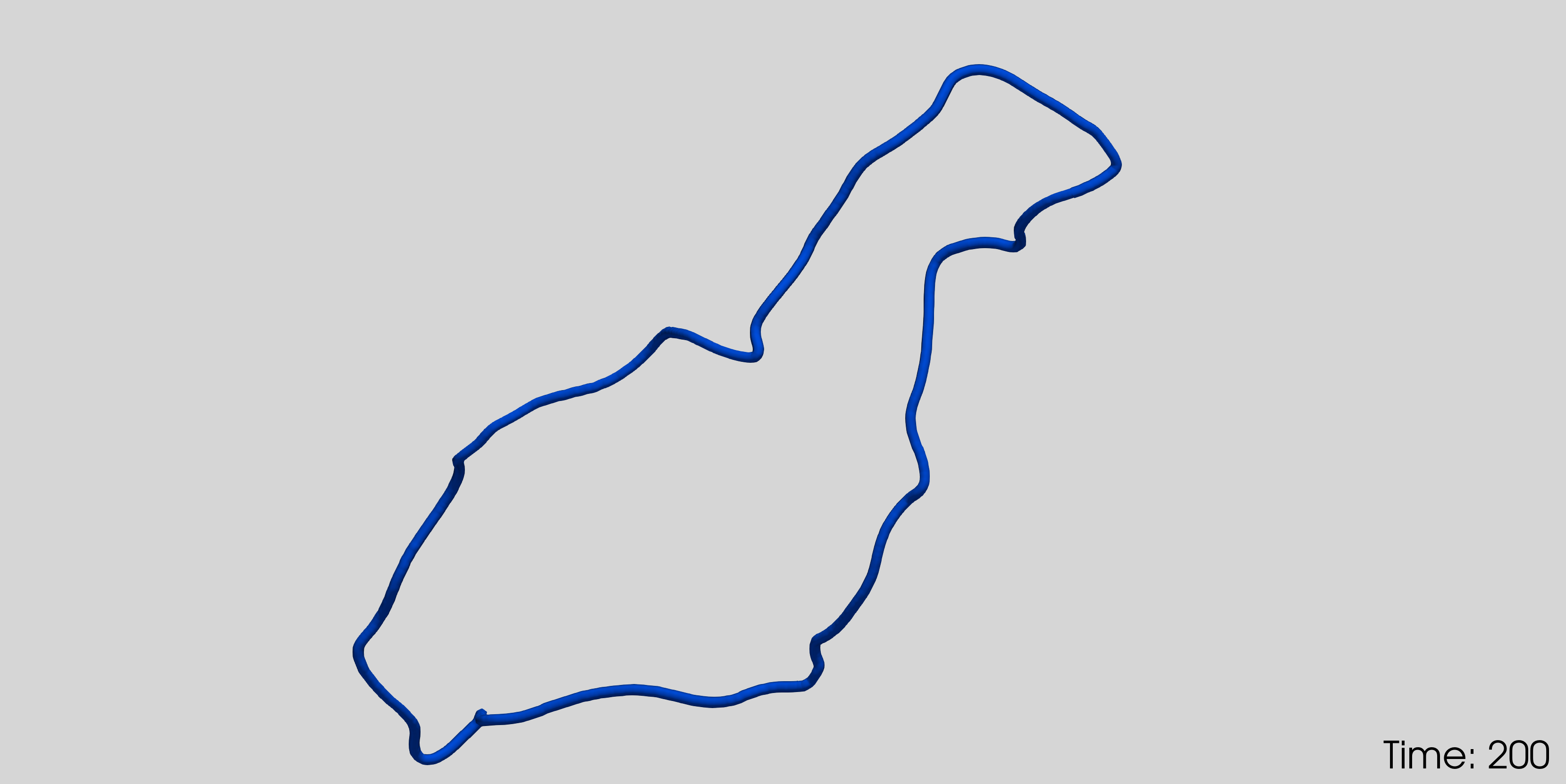}\vspace{2pt}
\\
\includegraphics[width=11cm, height=5cm]{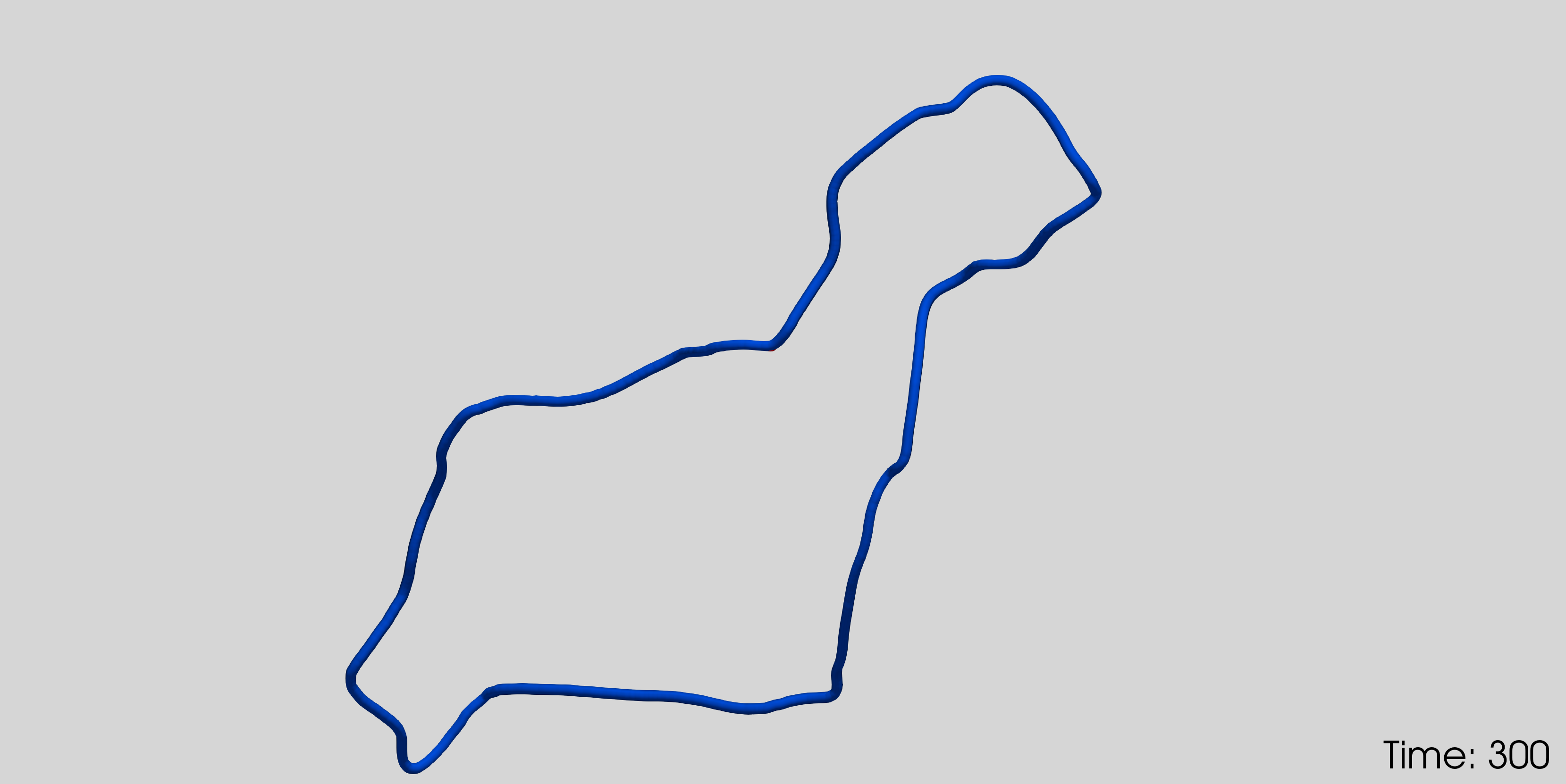}\vspace{2pt}\\
\includegraphics[width=11cm, height=5cm]{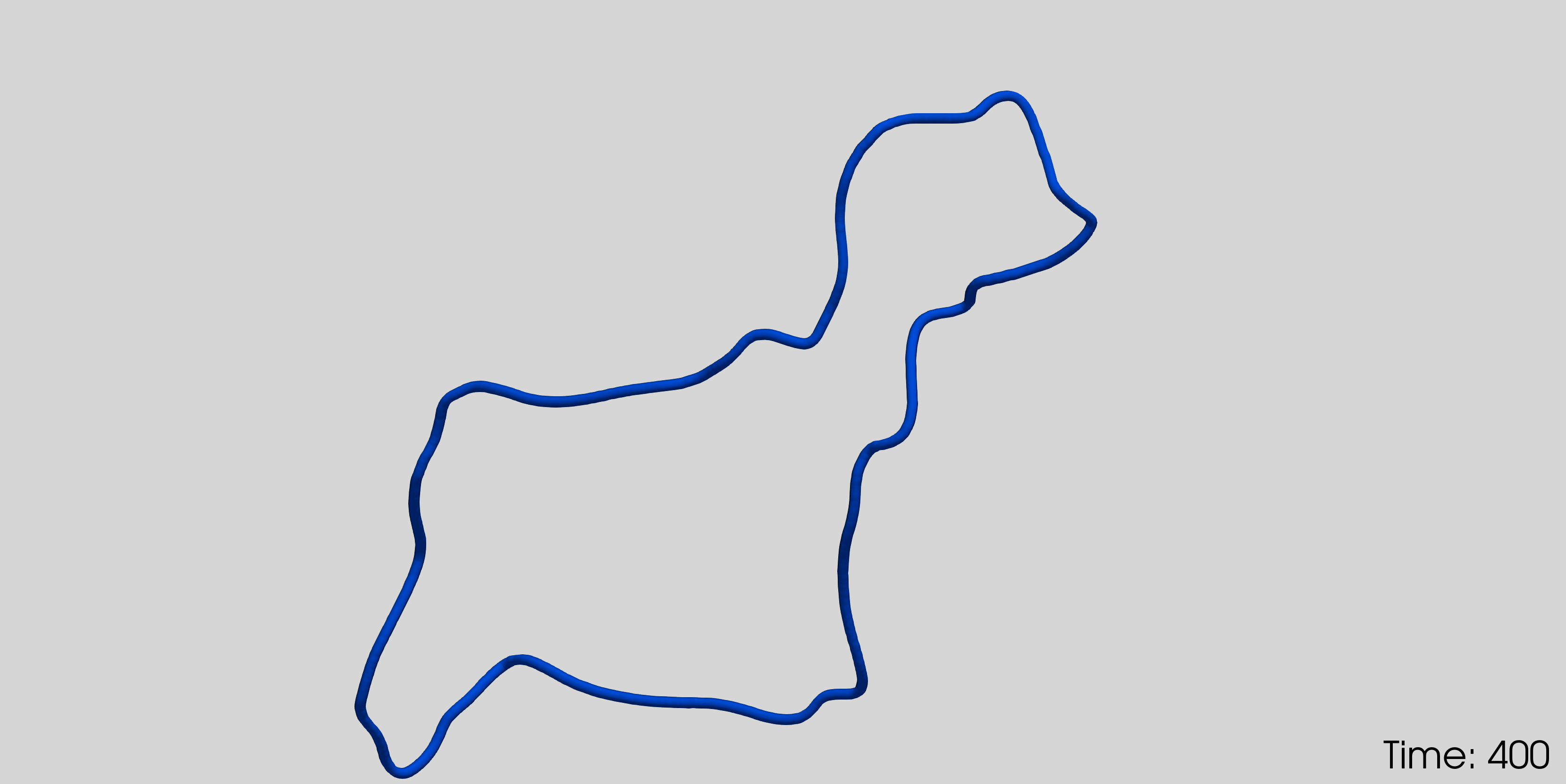}
\caption{The evolution of the field theory string seems to follow the NG 
prediction obtained from the reconstruction of the string after the high 
curvature event. There is no visible departure between the NG and the FT descriptions.}
\label{fig:after-reconstruction}
\end{center}
\end{figure}

In order to do a better job of identifying the reason for the
different evolution we follow a procedure also used in the past in
\cite{Olum:1998ag}. After we have identified one of these high
curvature events on the string, we reconstruct the NG data again. This
yields different ${\bf a}$ and ${\bf b}$ functions that should be
valid for the subsequent evolution. The interesting point is that the
field theory after these events have passed is again accurately
described by the new NG data. We show in
figure~\ref{fig:after-reconstruction}
a comparison of the position of the string obtained from 
field theory to the NG reconstruction after the first
episode of high curvature radiation. We notice that there
is no visible departure of the field theory evolution from the
prediction of NG. 

\begin{figure}
\begin{center}
\includegraphics[width=10cm, height=5cm]{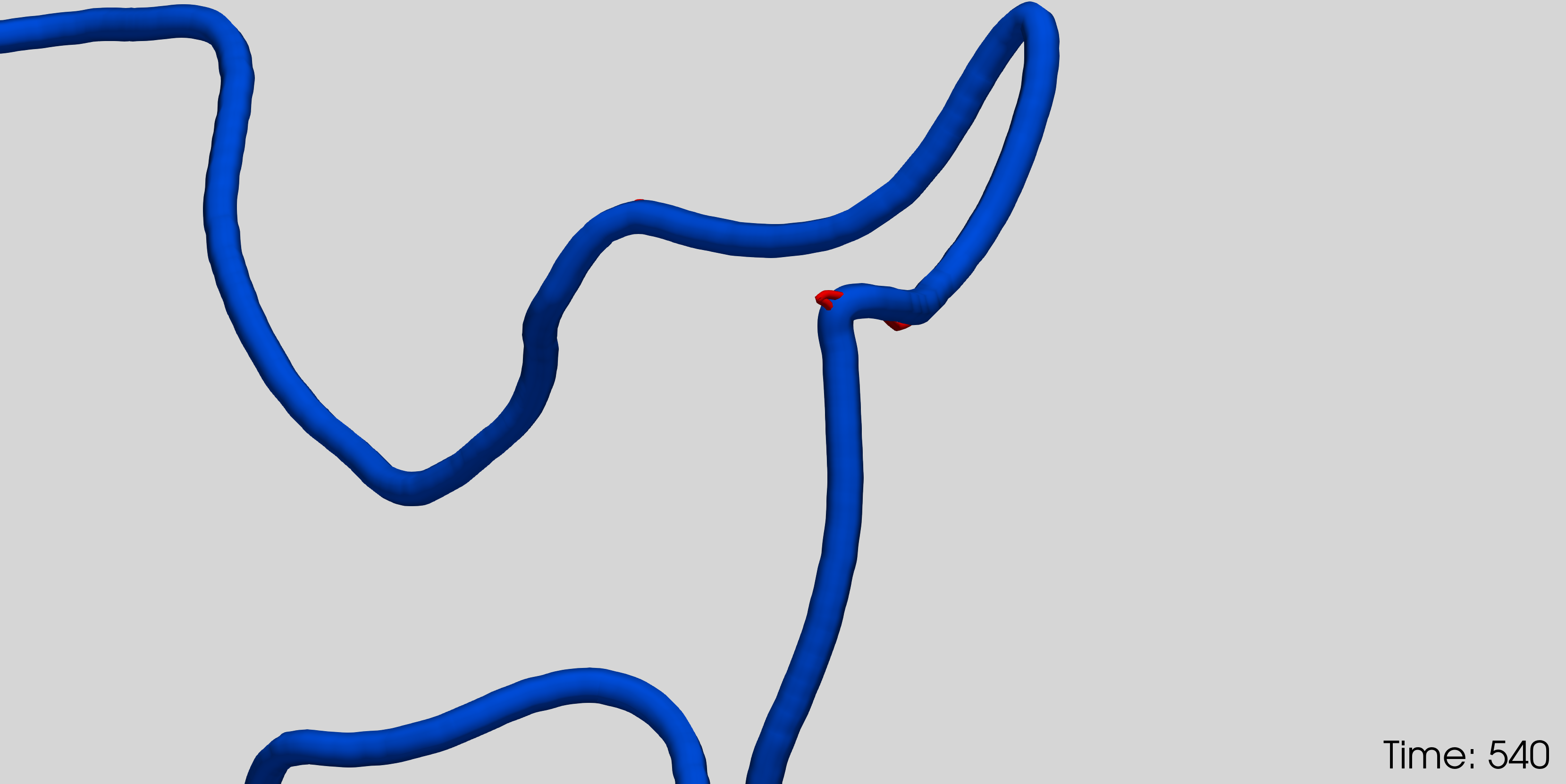}\vspace{2pt}\\
\includegraphics[width=10cm, height=5cm]{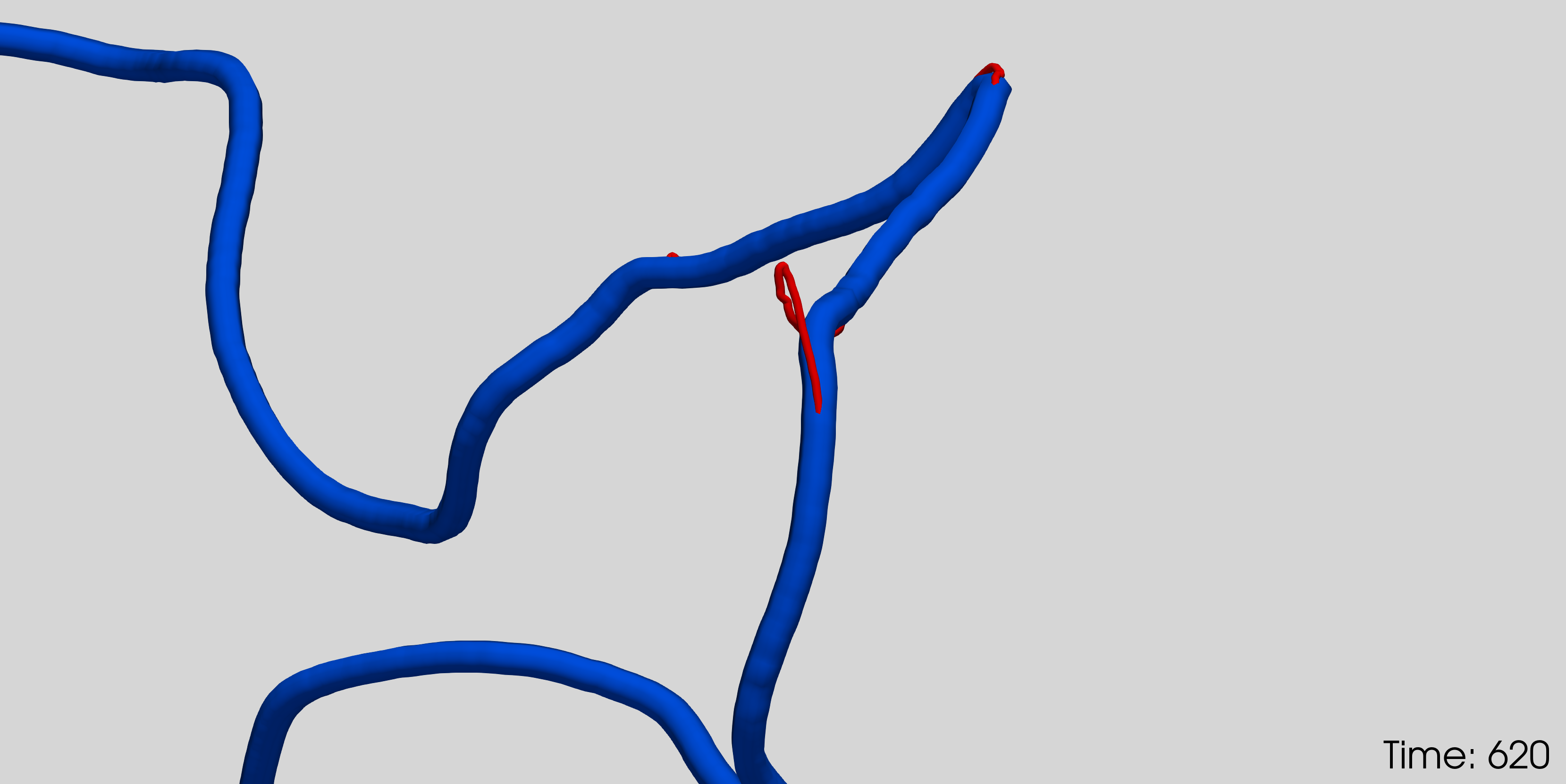}\vspace{2pt}
\\
\includegraphics[width=10cm, height=5cm]{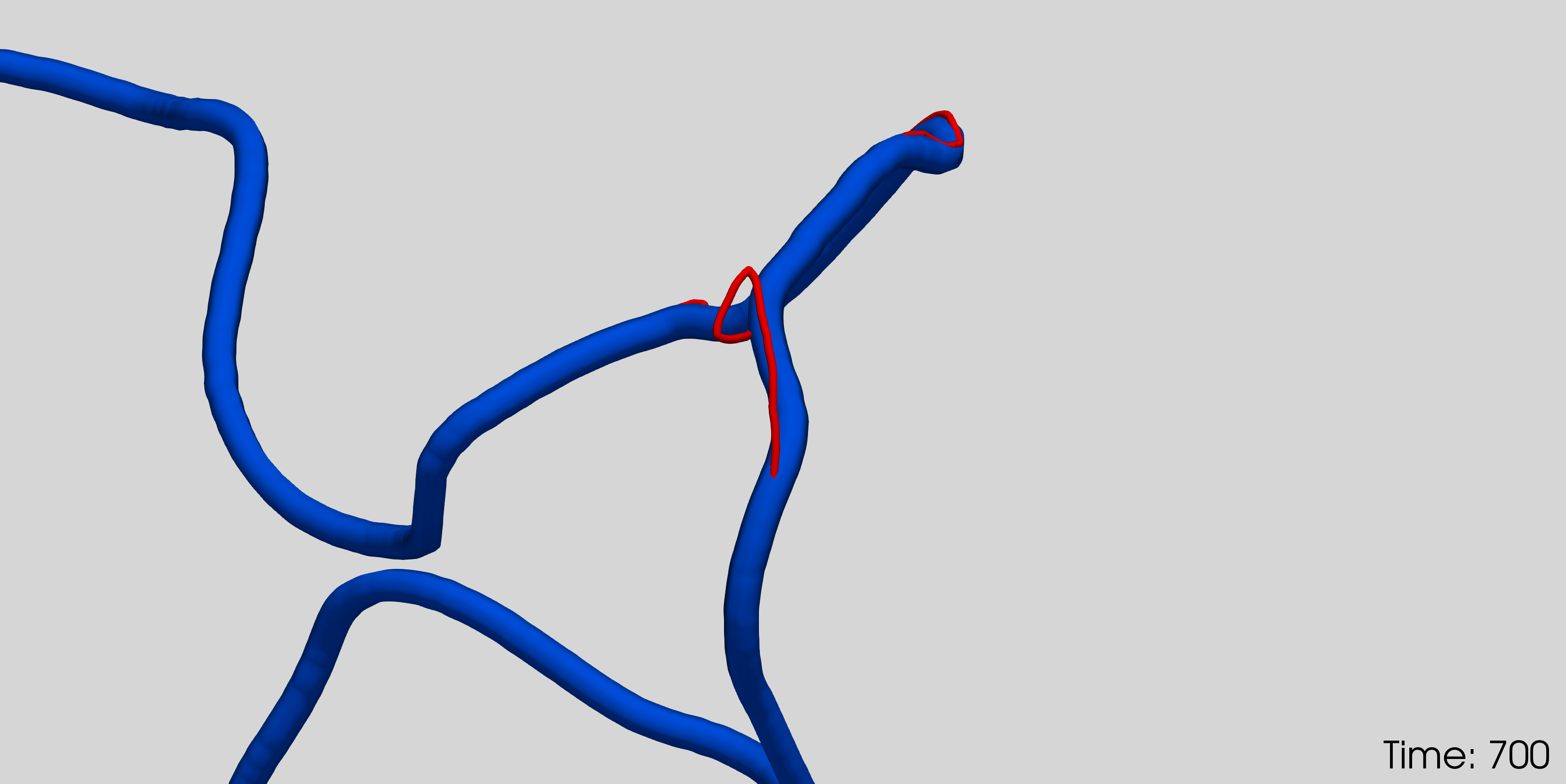}
\caption{Another high curvature event on the same loop. The NG description (in red)
is the one obtained after the first event in figure~\ref{fig:first-event}.}
\label{fig:second-event}
\end{center}
\end{figure}

This behaviour continues for a while until a new episode occurs. This is
shown in figure~\ref{fig:second-event},
where we clearly see another region where the NG prediction deviates
from the field theory result in a localized region. It is clear that
the string in field theory does not want to curve itself so much as
the NG predicts and takes a shortcut.  This process radiates again
some portion of the energy of the string.

\begin{figure}
\begin{center}
\includegraphics[width=10cm, height=5cm]{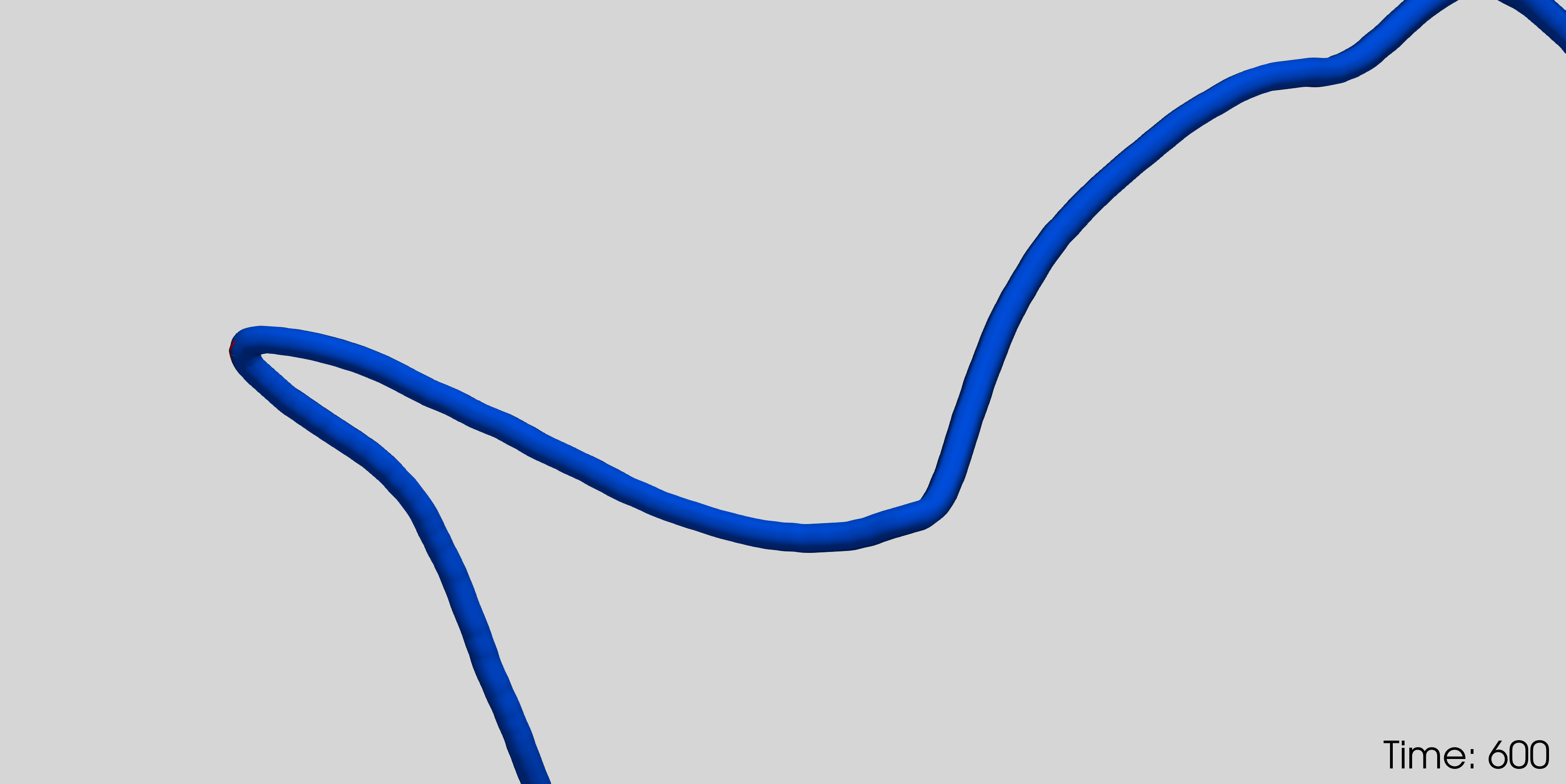}\vspace{2pt}
\\
\includegraphics[width=10cm, height=5cm]{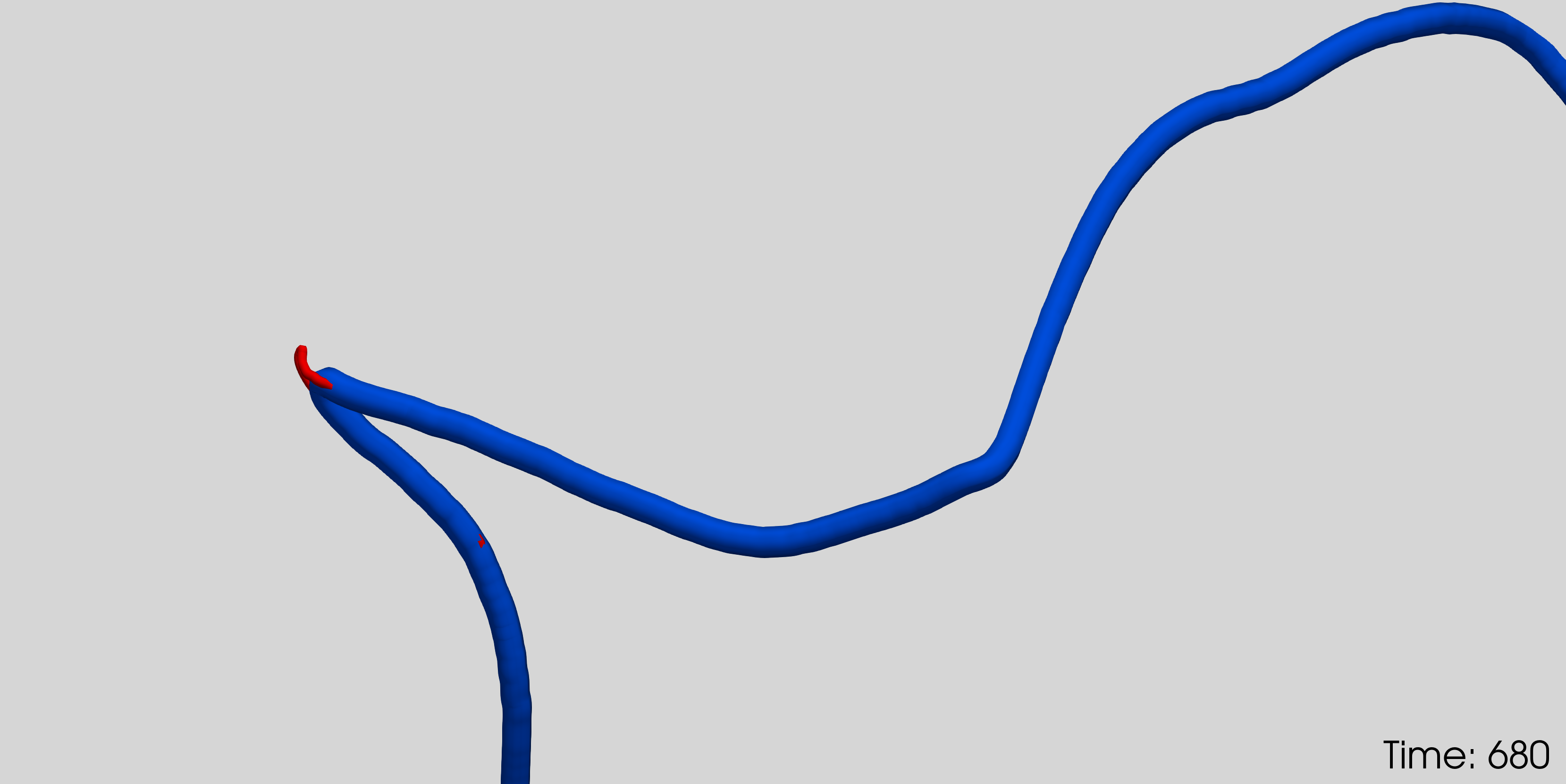}\vspace{2pt}
\\
\includegraphics[width=10cm, height=5cm]{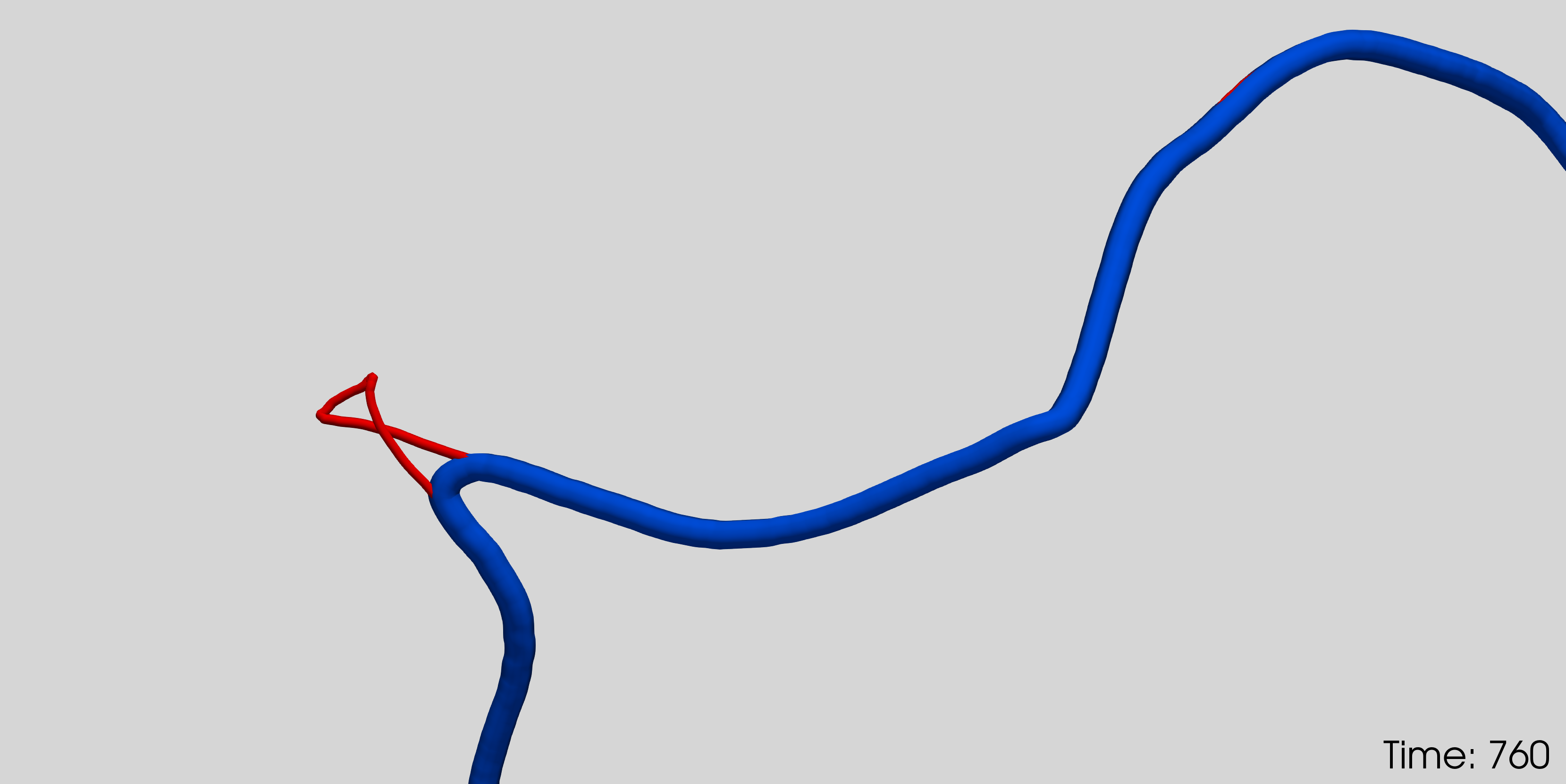}
\caption{Another high curvature event on the same loop. The NG description (in red)
is the one obtained after the first event in figure~\ref{fig:first-event}. The evolution between
these episodes is well represented by the NG dynamics.}
\label{fig:third-event}
\end{center}
\end{figure}

We have seen a similar behaviour in all our loops. Some of the
examples are clear, but some other ones are harder to visualize
since more than one of these high curvature events happen to have some
non-trivial overlap in time \footnote{However, we would like to emphasize that 
the results we present here, with this particular loop, are indeed a good 
representation of what we obtain in our analysis of all the other $6$ loops from \cite{Hindmarsh:2021mnl}. }.
  In fact, this already happens in our
example loop.  We show in figure~\ref{fig:third-event}
a third event situated quite far away in space from the previous one
but that overlaps in time with the event represented in
figure~\ref{fig:second-event}.

\begin{figure}
\begin{center}
\includegraphics[width=12cm, height=8cm]{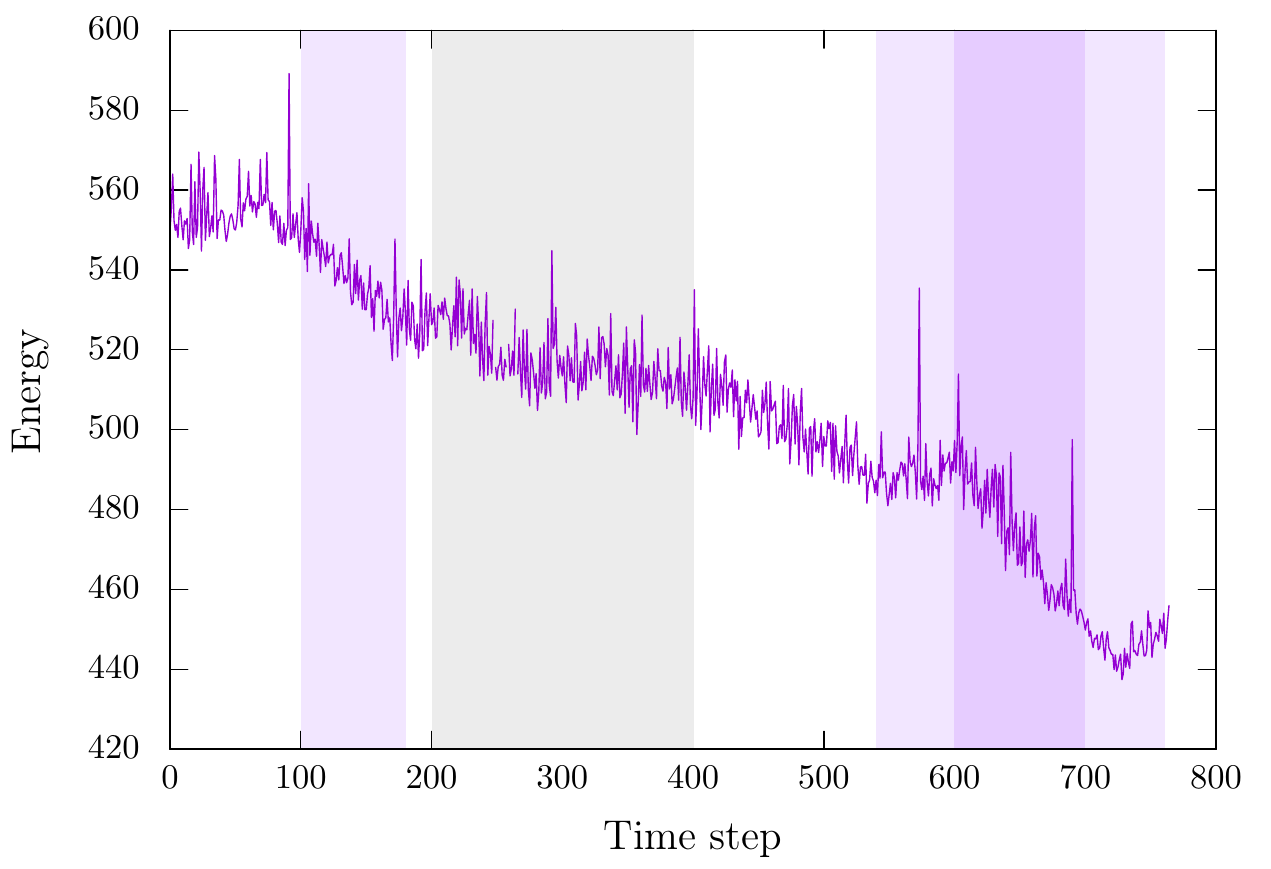}
\caption{ Total amount of invariant energy (total amount of $\sigma$) for the
NG reconstruction of the field theory data of the loop analyzed in other figures. The purple regions correspond to the high curvature events presented in figures~\ref{fig:first-event},\ref{fig:second-event} and \ref{fig:third-event}. The grey region corresponds to figure~ \ref{fig:after-reconstruction}, where the energy remains roughly constant. The white region (between $t\sim 400$ and $550$) corresponds to 
figure~\ref{fig:last}, where there is a small deviation from the NG dynamics.} 
\label{fig:NG-energy-reconstructed}
\end{center}
\end{figure}

We also look at the energy computed from the local
reconstruction of the NG string obtained from the field theory 
data at any moment in time. In a purely NG dynamics, this quantity (the total amount
of $\sigma$ of the loop) should be a constant of motion.
We plot in figure~\ref{fig:NG-energy-reconstructed} this energy for the same loop that we
discussed before. We observe that the energy overall tends to go down. There are some episodic events where the decrease in energy is sharper, and some of those can be linked with the high-curvature events. We mark in figure~\ref{fig:NG-energy-reconstructed}
the three different episodes that we have been discussing above by
shading in light purple the ranges of times displayed in
figures~\ref{fig:first-event}, \ref{fig:second-event}, and
\ref{fig:third-event}.  Looking at the energy, there seems to be a
connection between these events and the regions where the energy
starts to decrease.

There are other instances (between $t\sim200$ and $t\sim400$) where the energy seems to be constant, represented in grey in figure~\ref{fig:NG-energy-reconstructed}. These correspond to the times where the NG reconstruction was made after the first episode, as shown in figure~\ref{fig:after-reconstruction}, where there is a very good agreement with the FT data.

\begin{figure}
\begin{center}
\includegraphics[width=11cm, height=5cm]{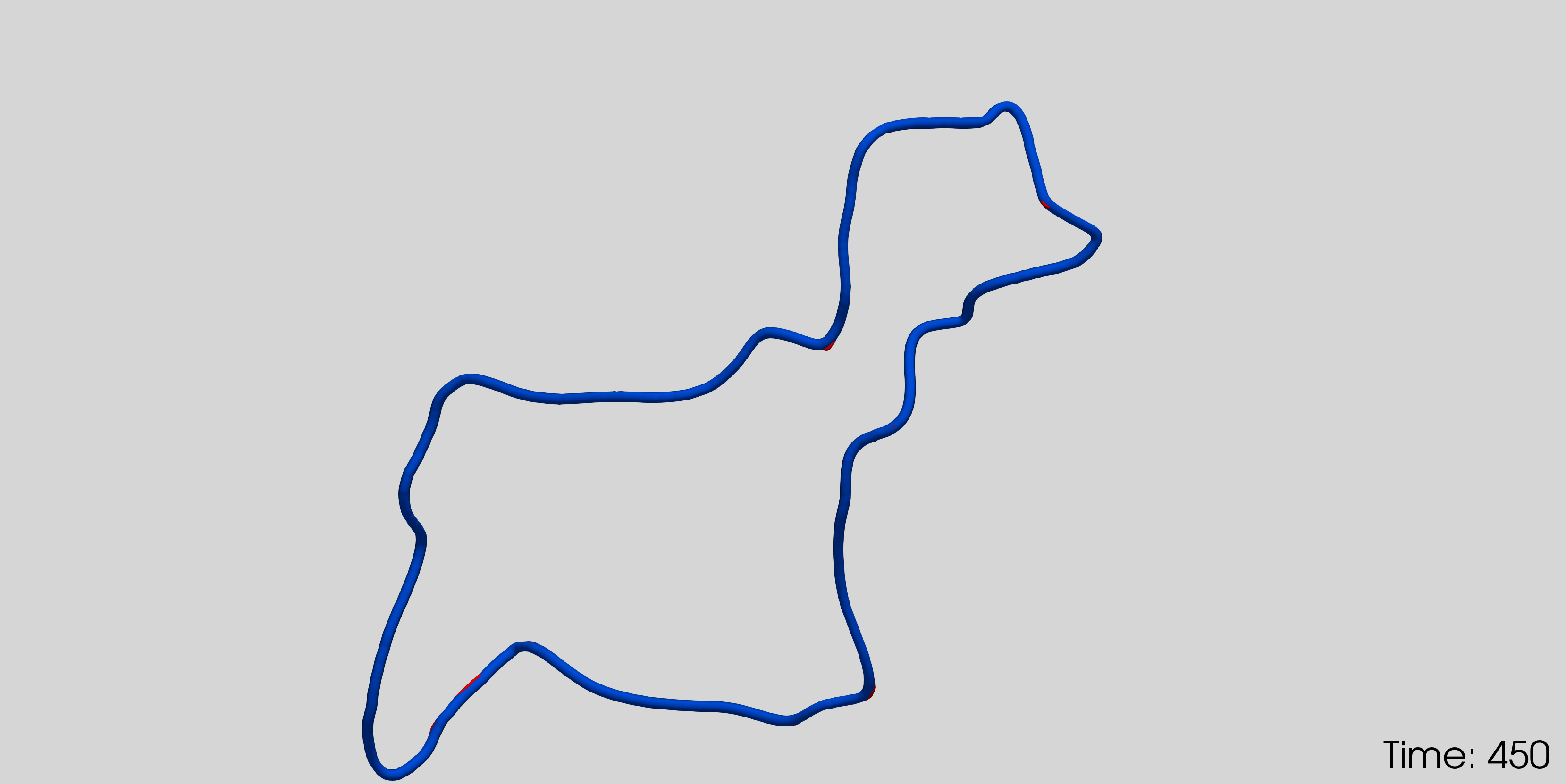}\vspace{2pt}
\\
\includegraphics[width=11cm, height=5cm]{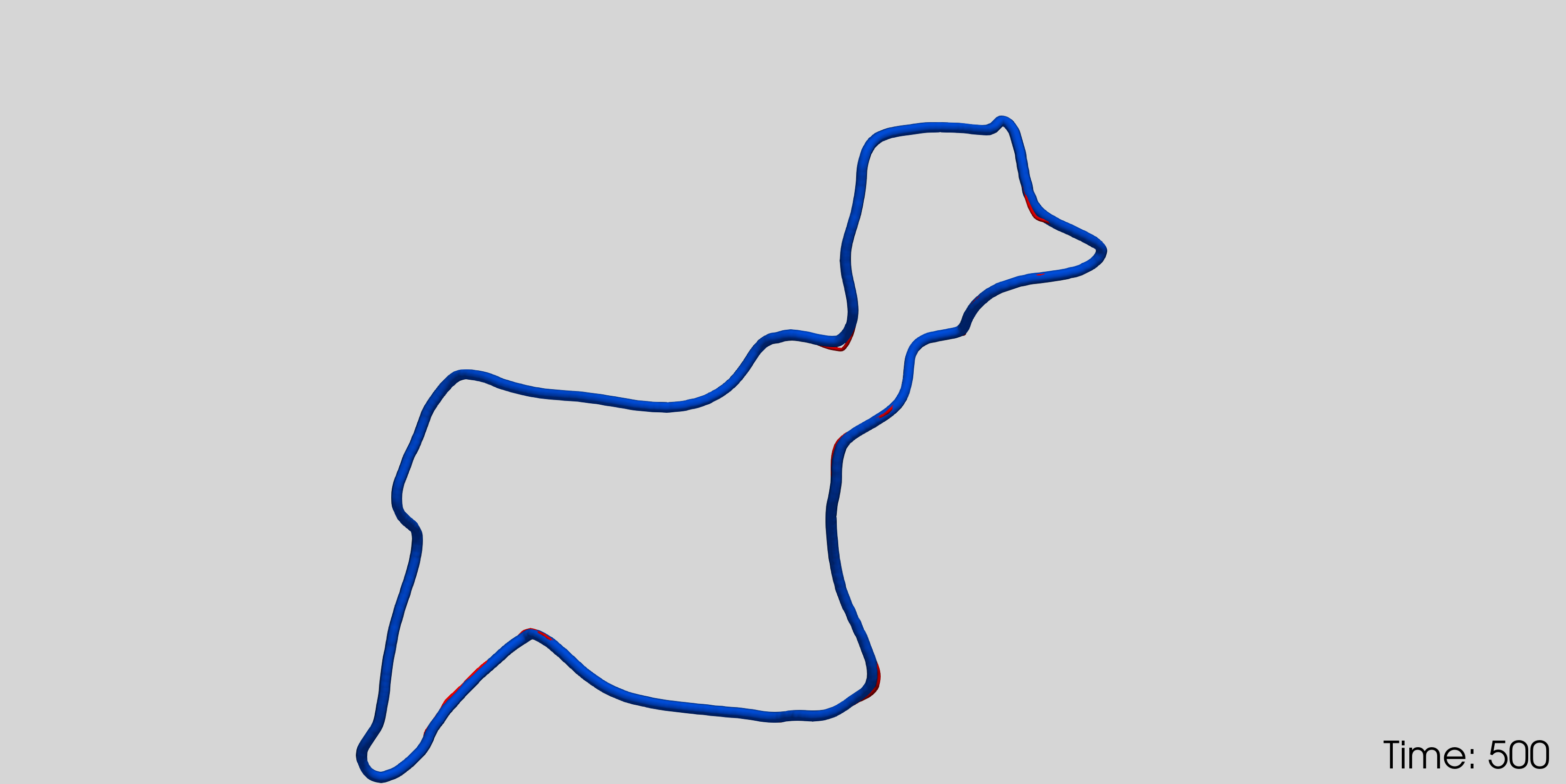}\vspace{2pt}\\
\includegraphics[width=11cm, height=5cm]{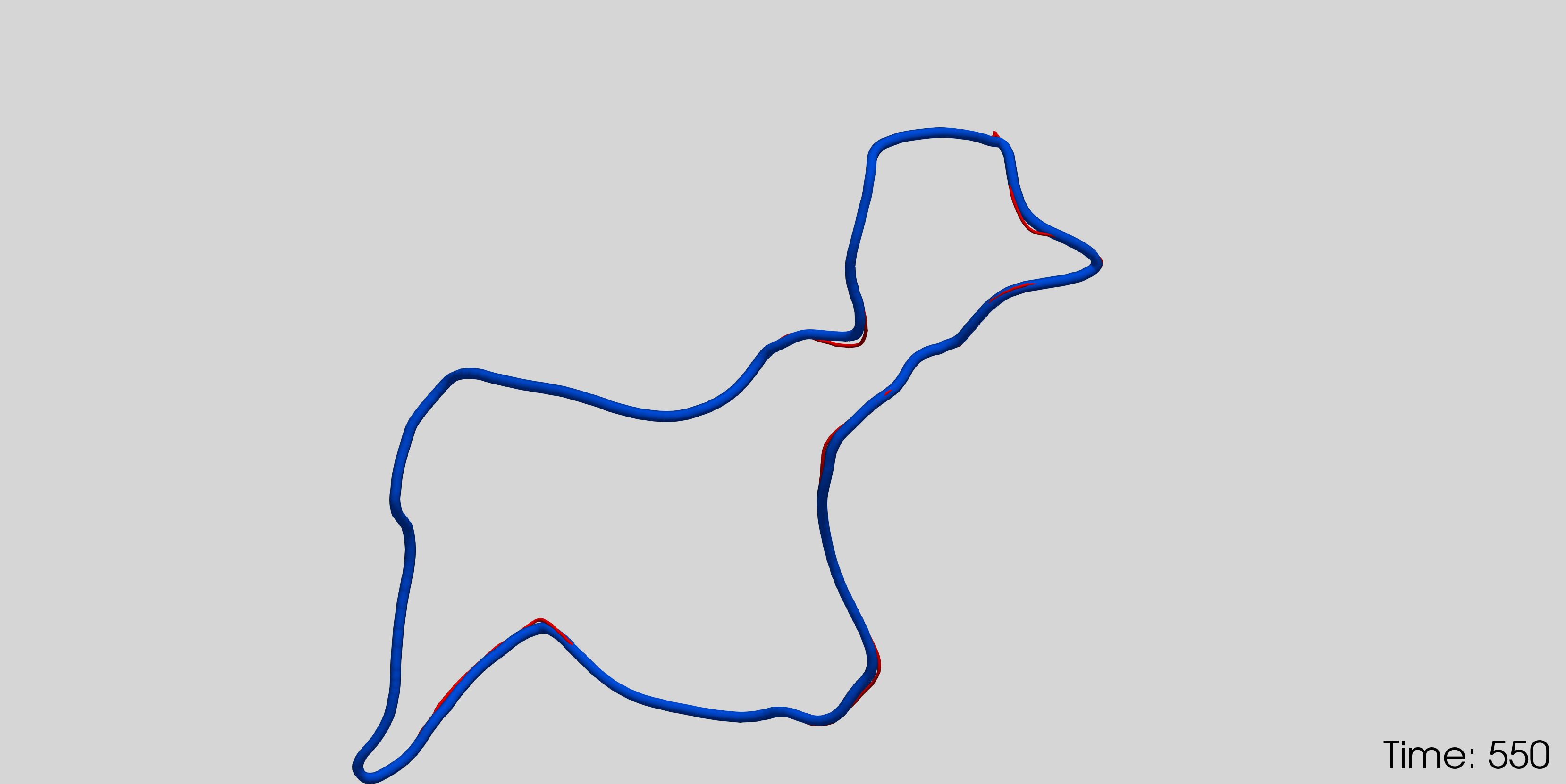}
\caption{The evolution of the field theory string in times where the energy decreases. One can not identify any single event of high curvature
associated with this decline in energy. Instead we see several regions of high curvature where the deviation from NG is clearly visible but not as
striking as in the previous examples.}
\label{fig:last}
\end{center}
\end{figure}

There are, however, regions where the energy slowly
decreases that are not so obviously associated with any of these
individual events (see figure~\ref{fig:last}, which corresponds to times 450, 500 and 550). The reasons for these are not so clear. The curvature does not seem to be very high, 
and could also be understood as small scale structure in the loop that leads to energy loss by radiation. In other words, this could be due to events similar to the ones
already mentioned that are not so  clearly visible in our procedure. Looking at figure~\ref{fig:last} one indeed sees several regions of slight departure
between NG and FT dynamics, especially at $t=550$.
 Note that identifying small
regions of high curvature from the NG reconstruction is sometimes
quite difficult due to numerical error. In the future we will 
design new methods to quantify this effect and understand
 the reason for this energy loss in these regions as well. It is remarkable, and somewhat puzzling, that though the energy goes down 
 roughly by the same percentage as in a high-curvature event, the visual inspection of the loop dynamics does show a rather small 
 deviation from the NG trajectory. This could just be due to
 the combined effect of several smaller regions where the deviation is small instead of a single large event like in the other cases.

\section{Conclusions}
\label{sec:conclusion}

In this paper we have compared the evolution of field theory loops
obtained in the course of a cosmological lattice simulation with their
expected dynamics in the NG approximation.  Understanding the
discrepancy between these two approaches is of paramount importance in
order to make an accurate prediction of the observational signatures
of strings. In particular, it is crucial in the estimate of the
gravitational wave signature from strings in current and future
gravitational wave observatories (see \cite{Blanco-Pillado:2017oxo}
and references therein).

Our investigations show that loops in field theory seem to
behave according to the NG action in regions where the curvature is not high. The visual comparison with the NG motion 
does not support the need for any departure in the equation
of state of field theory loops, also, in regions where the curvature is not high. The strings move with the local
trajectories dictated by NG. However, we have found that
the strings lose part of their energy in the
course of their evolution. Some of these energy loss regions correspond to 
high curvature sections where some portion of the string 
annihilates with an adjacent part. These high curvature events 
are indeed predicted by the NG evolution obtained from the original
reconstruction of the field theory string. Of course, the NG dynamics cannot
account accurately for the subsequent evolution of the
field theory string since part of the energy of the string
is lost in these events. This can be bypassed by reconstructing
the NG data again after one of these incidents. The result we obtain
by following this prescription seems to show that the evolution of the strings
is again described by NG with this new data.
There are other instances where the string loses energy, which cannot be so clearly pinpointed as regions of high curvature, at least following our visual inspection. Remarkably, though, the trajectory of the string does not seem to be altered perceptibly in these events. They deserve further investigation.

The picture that emerges from our detailed comparison of both
descriptions of the string motion is the following.  Most of the time,
loops behave as NG predicts, but there are instances where the NG
action breaks down and one needs to interrupt this comparison for a
while until the NG behaviour resumes again. The study of the conserved NG energy backs up that there are instances where the energy drops that correspond to high curvature events.

This localized energy loss mechanism makes the loops shrink and
sometimes self-intersect before they have a chance to oscillate for a
full period. So at the end the resultant loops are too small to expect them to behave
as NG and they finally disappear. This could explain why we do not get at the end any
non-self-intersecting loop from our simulation even though the dynamics of loops is well
explained by NG for most of their evolution.

The question arises then: if the loops behave almost everywhere like NG, then, would one expect to get NSI loops also in FT, and thus a big chunk of the energy of the network be released as GW? The direct obvious way of answering this question might be to keep simulating loops of this kind until a NSI loop is found in field theory. This is not a good strategy, because, as indicated previously, NSI conditions are not so easy to 
come by. Many large NSI loops are found in NG simulations \cite{Blanco-Pillado:2011egf}, because one simulates a much, much larger
volume with many more loops \cite{Blanco-Pillado:2011egf}. Unfortunately, we do 
not have the dynamic range in field theory to do such simulations.
Of course, we may be lucky and find one such loop in our
simulations after a large number of them. 

Another idea would be to start with a different set of loops. For
example, we could get loops from field theory simulations in the
radiation or matter era. These loops should be smoother and have a
greater chance to become non-self-intersecting.

One would also be tempted to look for larger initial loops. However,
even though it is important to have large loops, so their size is
large compared to their thickness, what we have seen in these
simulations is that this is not the most important fact. One can have
a very large loop with wiggles that lead to high curvature regions,
which would thus lose energy by this mechanism. 

The best scenario
would be to start with a large enough loop that radiates most of this
energy in the high curvature regions in its first few moments leaving
behind a smoother loop that now should behave mostly as NG (except
maybe for the presence of cusps). This expectation is based on the results obtained in field theory
simulations of the collisions of wiggles that lead to high curvature
regions \cite{Olum:1999sg,Blanco-Pillado:2022rad}. That is the
analogous situation to what we are seeing here in these loops, but in
long, infinite strings.  The results there ndicate that the radiation from
these events decreases quite fast after their first encounter. The
wiggles become milder, and their subsequent interaction is not so
violent. This argues for a period of smoothing of the loops of the
order of one oscillation time after which the loops would become quite
close to NG in all their evolution. 

If this is the correct view, it means
that loops created in a real cosmological network have a transient
period where they emit massive radiation from these highly curved
regions. After this initial stage (of the order of the period of the
loop) this effect should smooth out the loop and loops would behave as the NG action predicts. Furthermore, as the universe
evolves, if the above picture is right, the sizes of structures on loops would increase proportionally 
to the horizon distance, while the string core size remains fixed. Thus, over 
cosmological time, the curvature radii seen on loops would become many 
orders of magnitude larger than the string thickness, the radiative processes 
we see here would disappear, and the loop motion would be accurately given by NG. 

Nevertheless, the study reported here is quite preliminary.  We have
analyzed only a few loops, and we do not understand their dynamics
completely.  So it is possible that more is going on than the simple
description above.  In that case many alternative scenarios
\cite{Vincent:1997cx,Hindmarsh:2008dw,Hindmarsh:2017qff,Hindmarsh:2021mnl} may be possible.

In summary, we believe we have made a major step forward in our
understanding of the dynamics of loops from field theory simulations.
It is clear that loops appear to move as NG for most of their evolution. One clear situation in which this is not happening is, not surprisingly, in events of high-curvature. There are other instances in which the energy of the loops does not seem to follow a NG prediction, and yet the behaviour of the loop does not present big departures from the NG trajectory. The details of this picture need to be confirmed by further study. In the future we will perform new numerical experiments and use different field theory simulations that could 
corroborate this picture. Some of this work is already underway.

\acknowledgments

We are grateful to Mark Hindmarsh for discussion and for letting us use the data from his work. We also thank Jose R.C.C.C. Correia, Carlos J. A. P. Martins, Jose M. Queiruga, Tanmay Vachaspati 
and Alex Vilenkin for useful discussions and collaboration on related work.  This study 
is supported in part by the PID2021-123703NB-C21 grant funded by MCIN/ AEI /10.13039/501100011033/ and 
by ERDF;  ''A way of making Europe'', the Basque Government grant (IT-1628-22), the Basque 
Foundation for Science (IKERBASQUE) and the National Science Foundation under grant number
2207267. Part of this work has been performed under the Project HPC- EUROPA3 (INFRAIA-2016-1-730897), 
with the support of the EC Research Innovation Action under the H2020 Programme; in particular, AU gratefully 
acknowledges the support of the Department of Physics of the University of Helsinki and the computer resources 
and technical support provided by CSC Finland, as well as support from the University of the Basque Country grant (PIF20/151).

\appendix

\section{Primordial loops}
\label{ploops}

As we explained in the main part of the text, we create our string
network with an initial period of diffusion. One can of course look at
the initial evolution of some of the loops created at this time as
well. However, we should note that they are quite different from the
ones we have analyzed in the rest of the paper. First of all, they are
much smoother due to the period of diffusion and furthermore they are
all created at rest, meaning all the segments of the string start
their evolution without any initial velocity. These properties make
these loops rather special from the point of view of their Nambu-Goto
evolution. It is easy to show that an initially static loop will
overlap with itself along the entirety of its physical length in a
half of its period if it moves according to the NG description \cite{Kibble:1982cb}.  It
is therefore clear that we cannot use these loops to illustrate the
typical behaviour of a loop in a realistic cosmological
setting. However, as we will describe in the following, we can use
these primordial loops to check the validity of our results and our
conclusions.

\begin{figure}
\begin{center}
\includegraphics[width=6.cm, height=4.2cm]{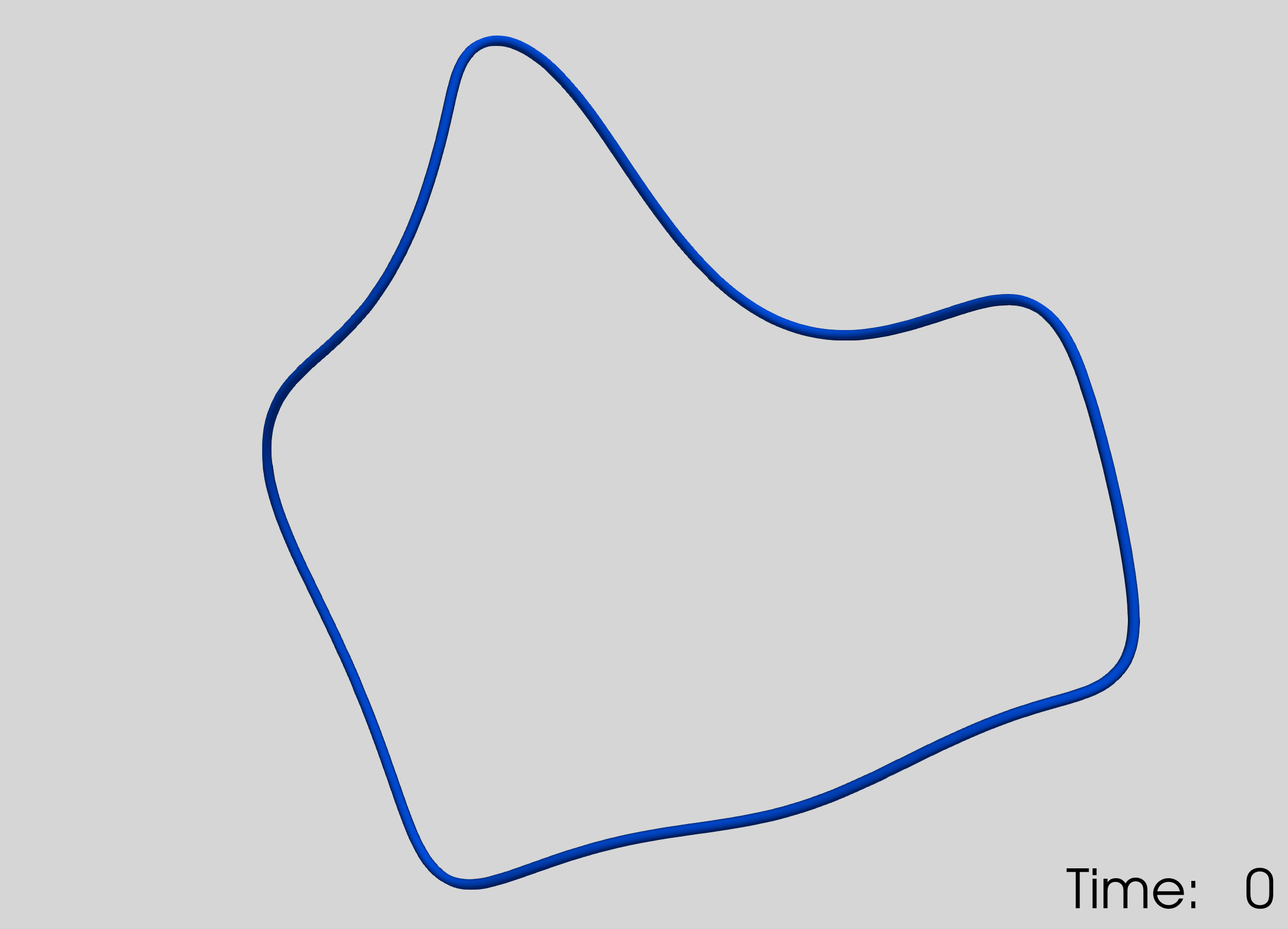}
\includegraphics[width=6.cm, height=4.2cm]{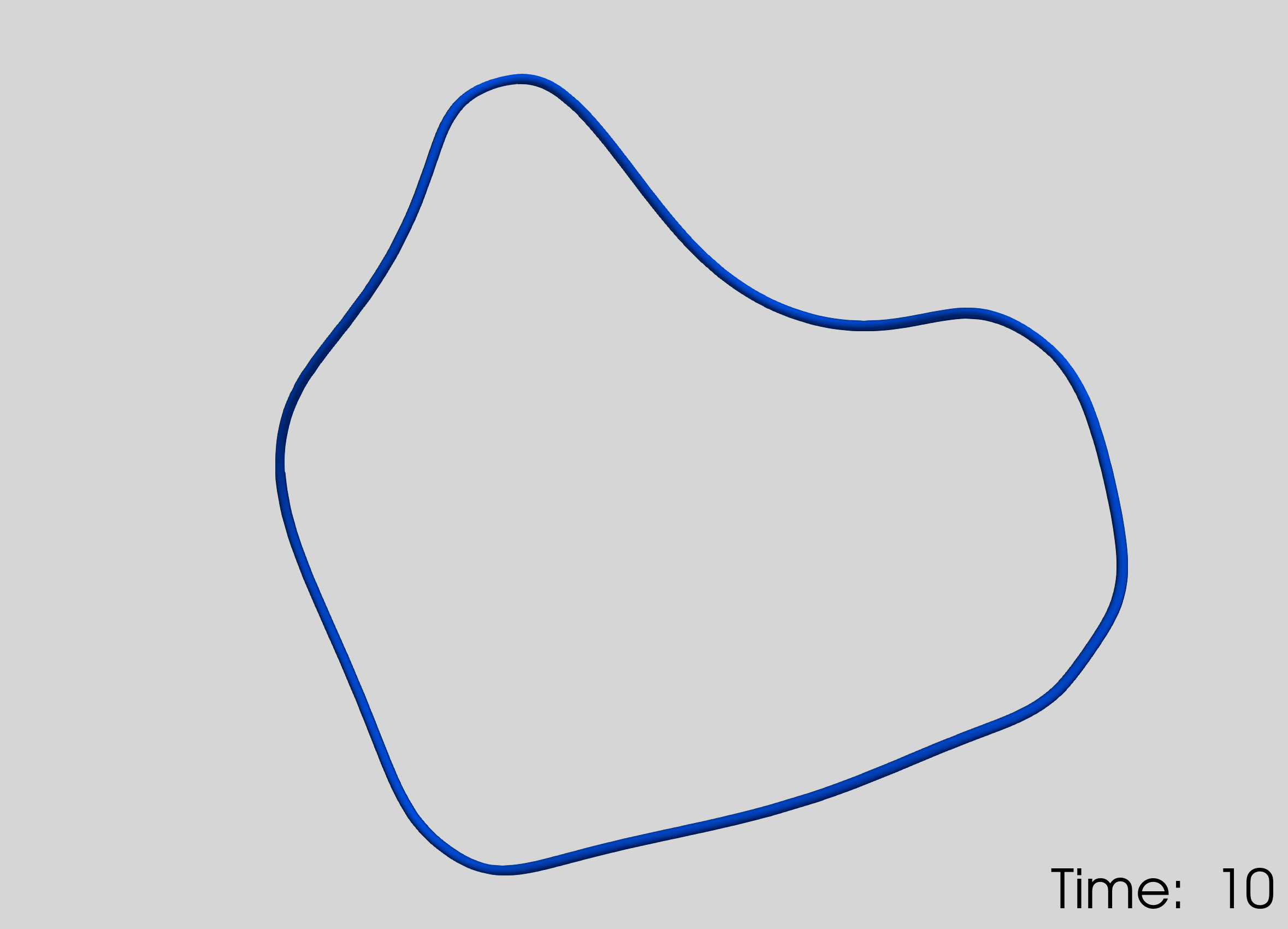}\vspace{2pt}\\
\includegraphics[width=6.cm, height=4.2cm]{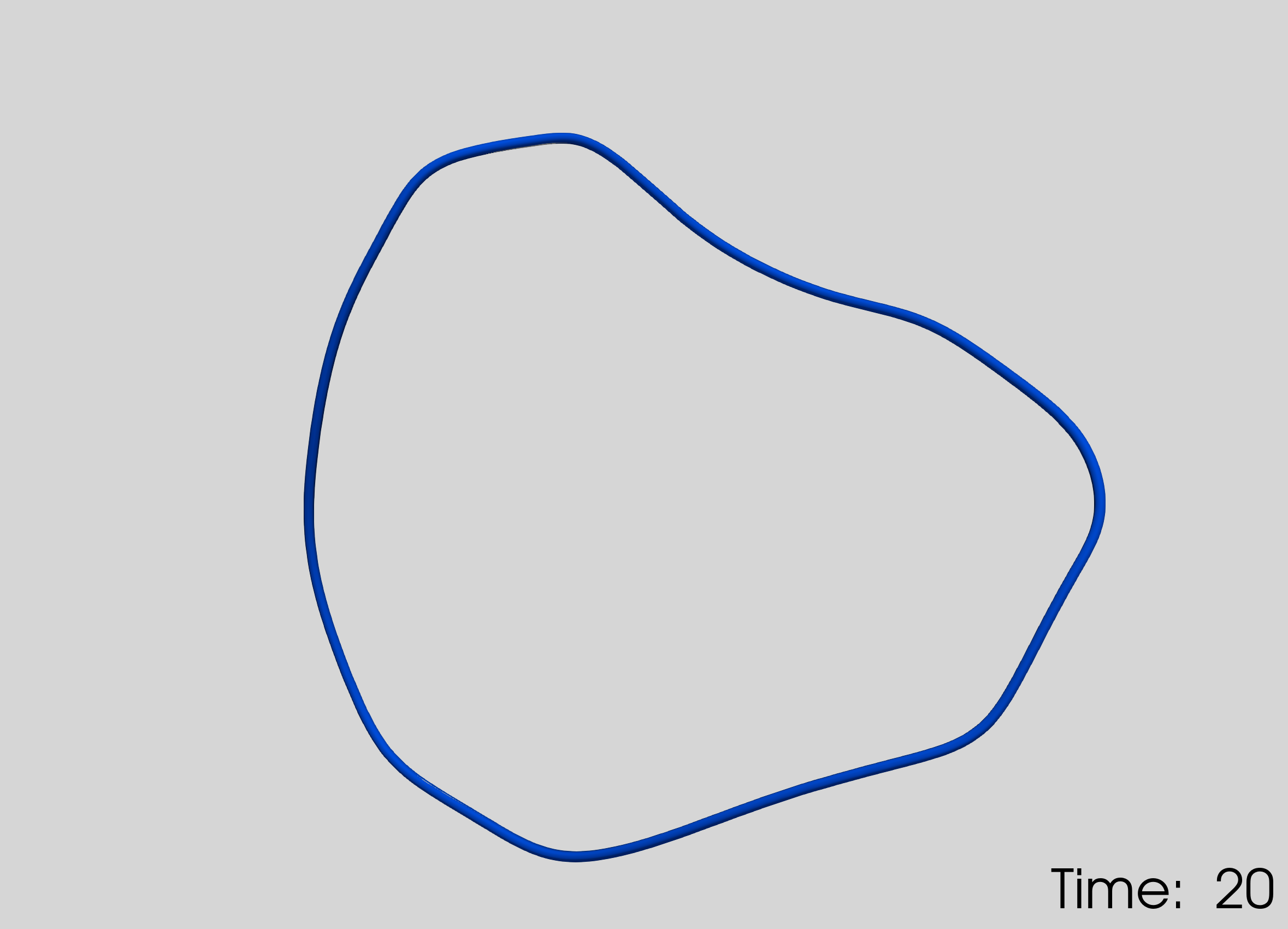}
\includegraphics[width=6.cm, height=4.2cm]{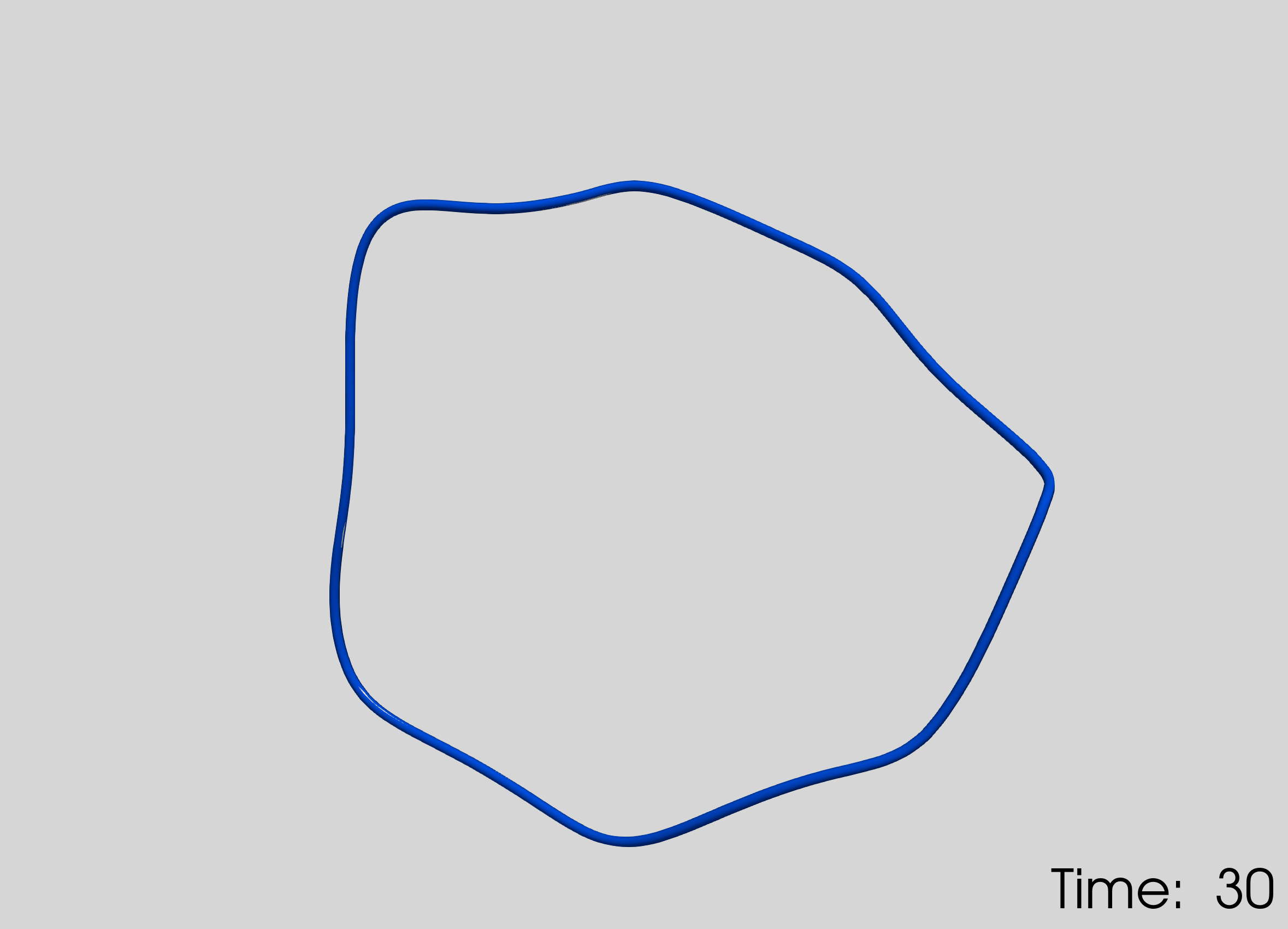}\vspace{2pt}\\
\includegraphics[width=6.cm, height=4.2cm]{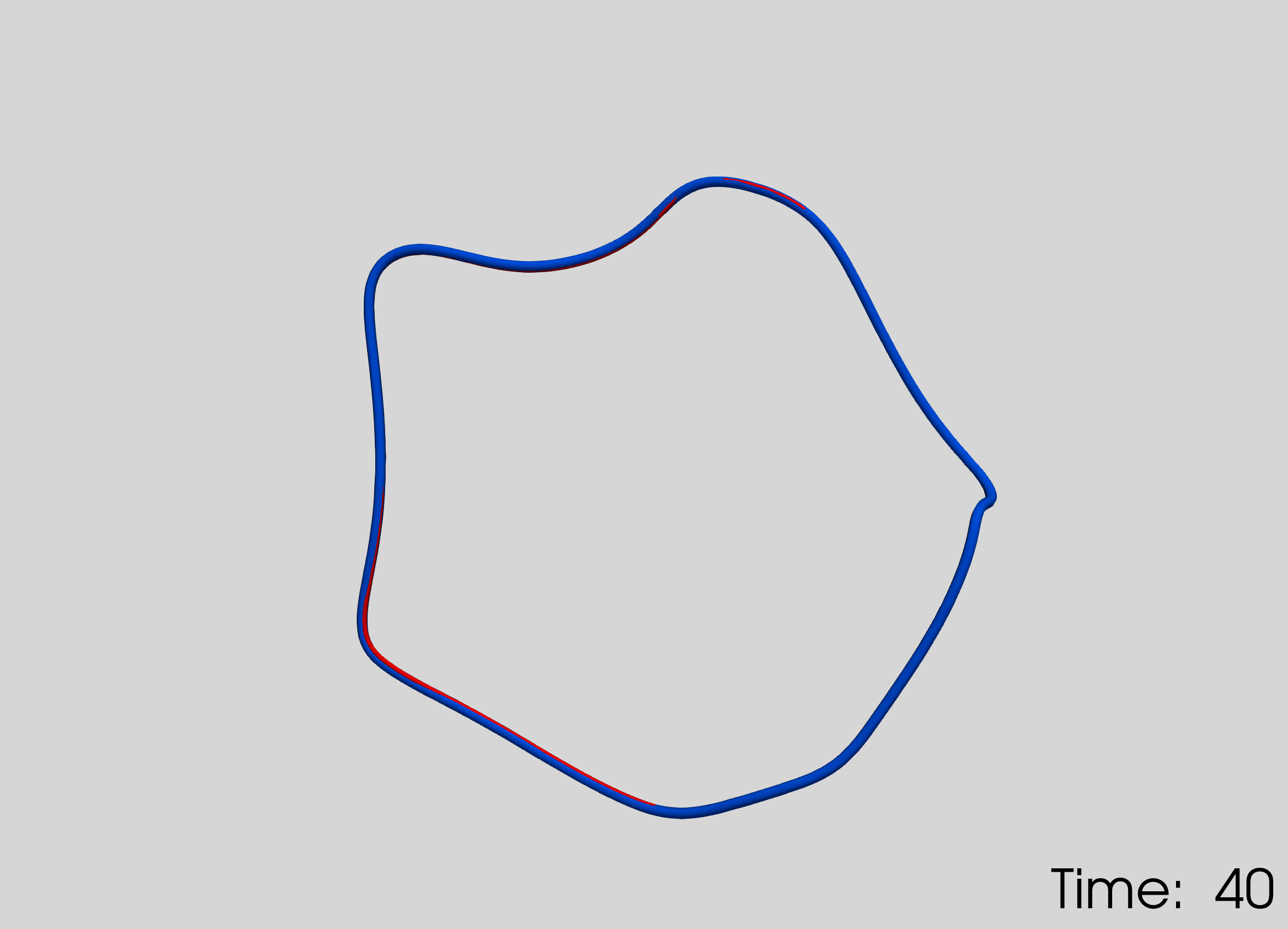}
\includegraphics[width=6.cm, height=4.2cm]{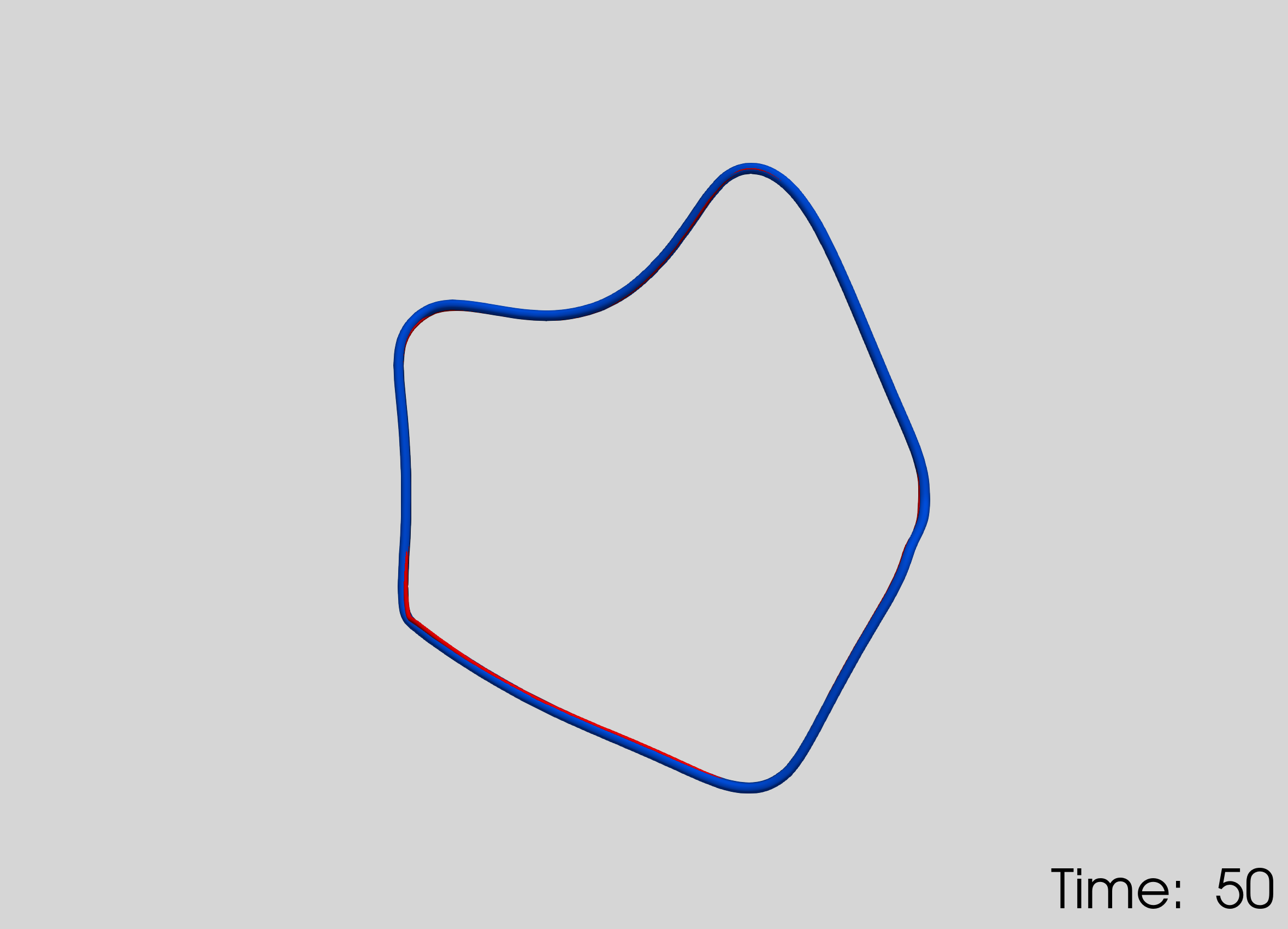}\vspace{2pt}\\
\includegraphics[width=6.cm, height=4.2cm]{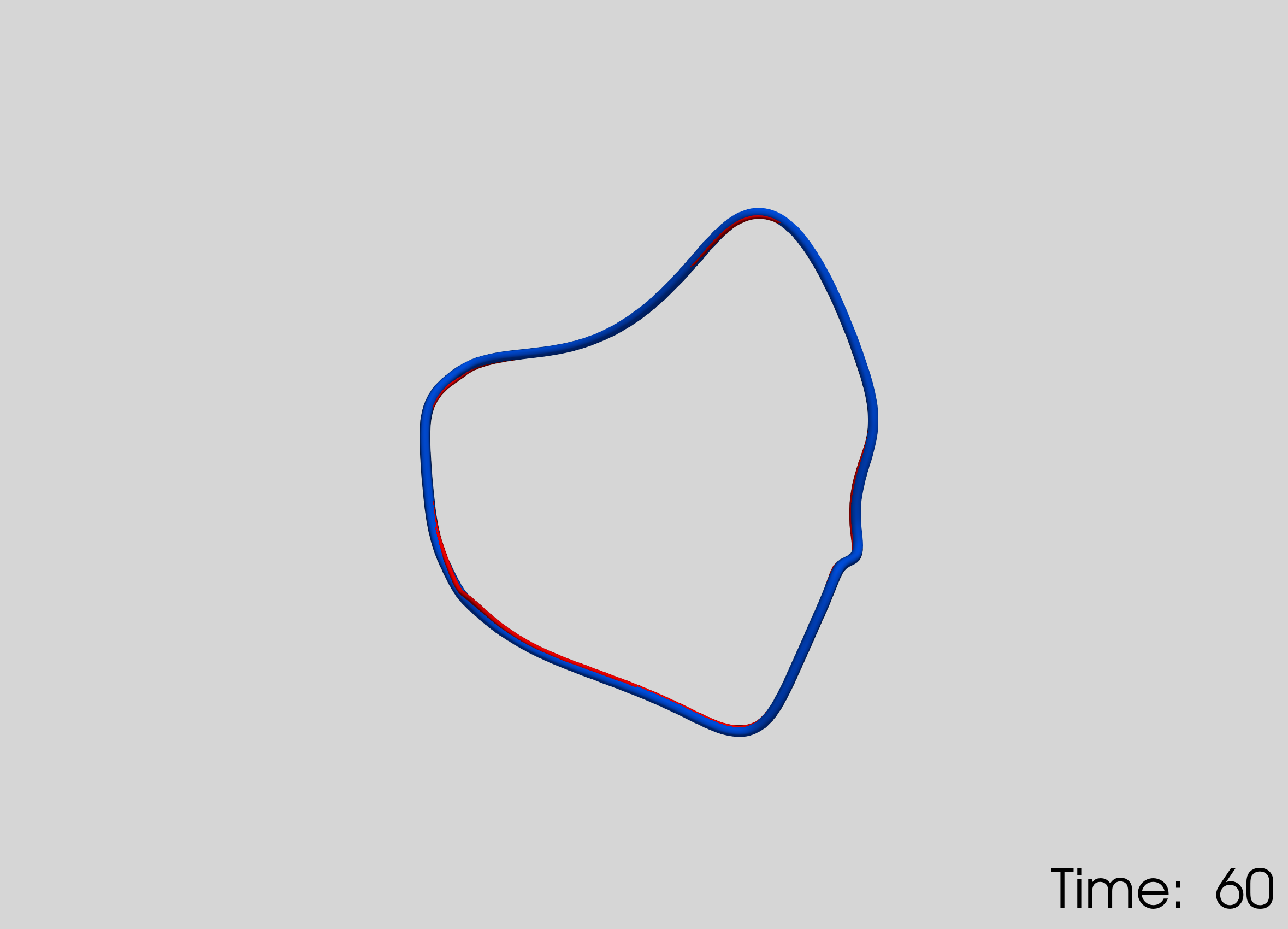}
\includegraphics[width=6.cm, height=4.2cm]{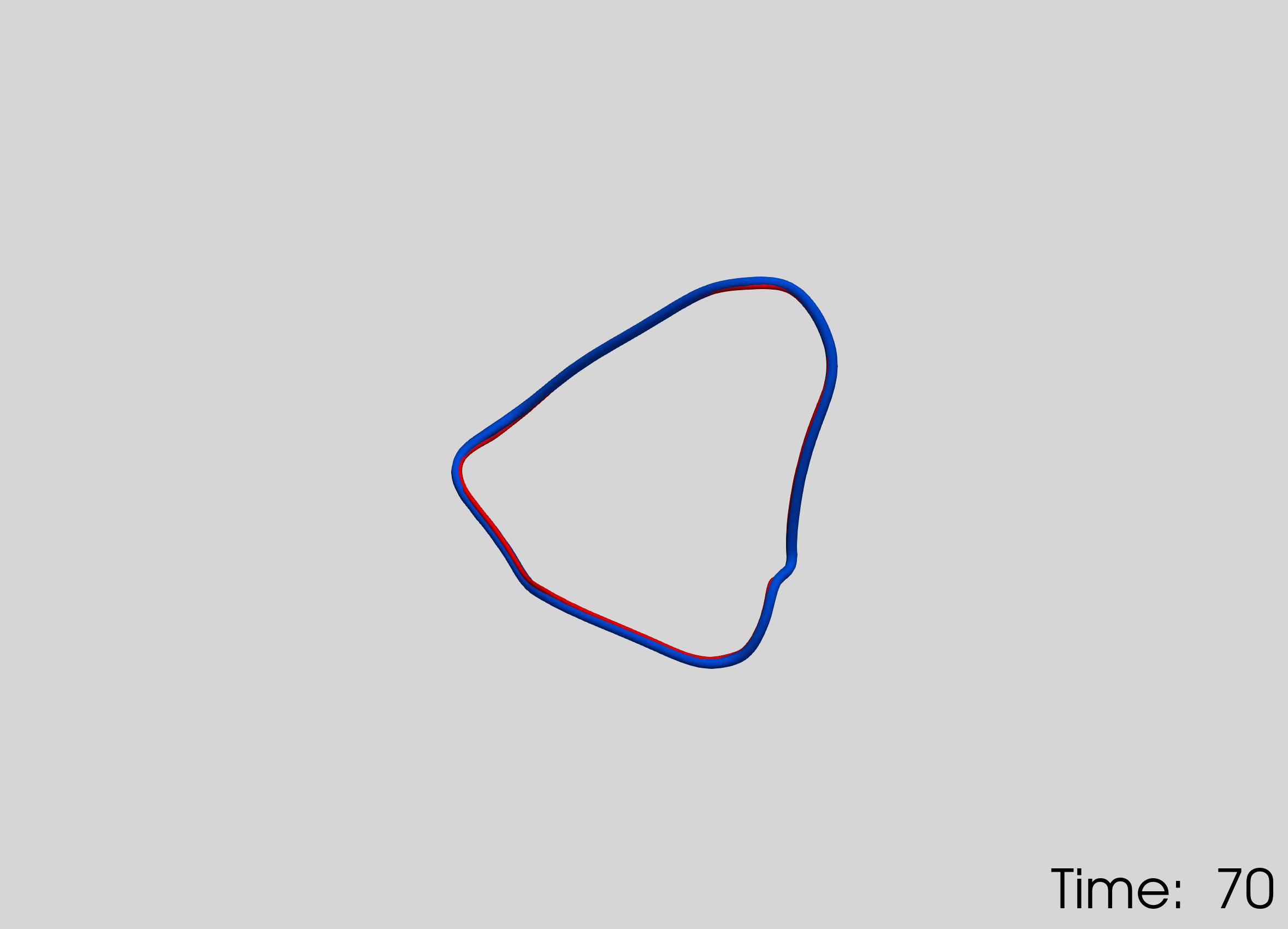}\vspace{2pt}\\
\includegraphics[width=6.cm, height=4.2cm]{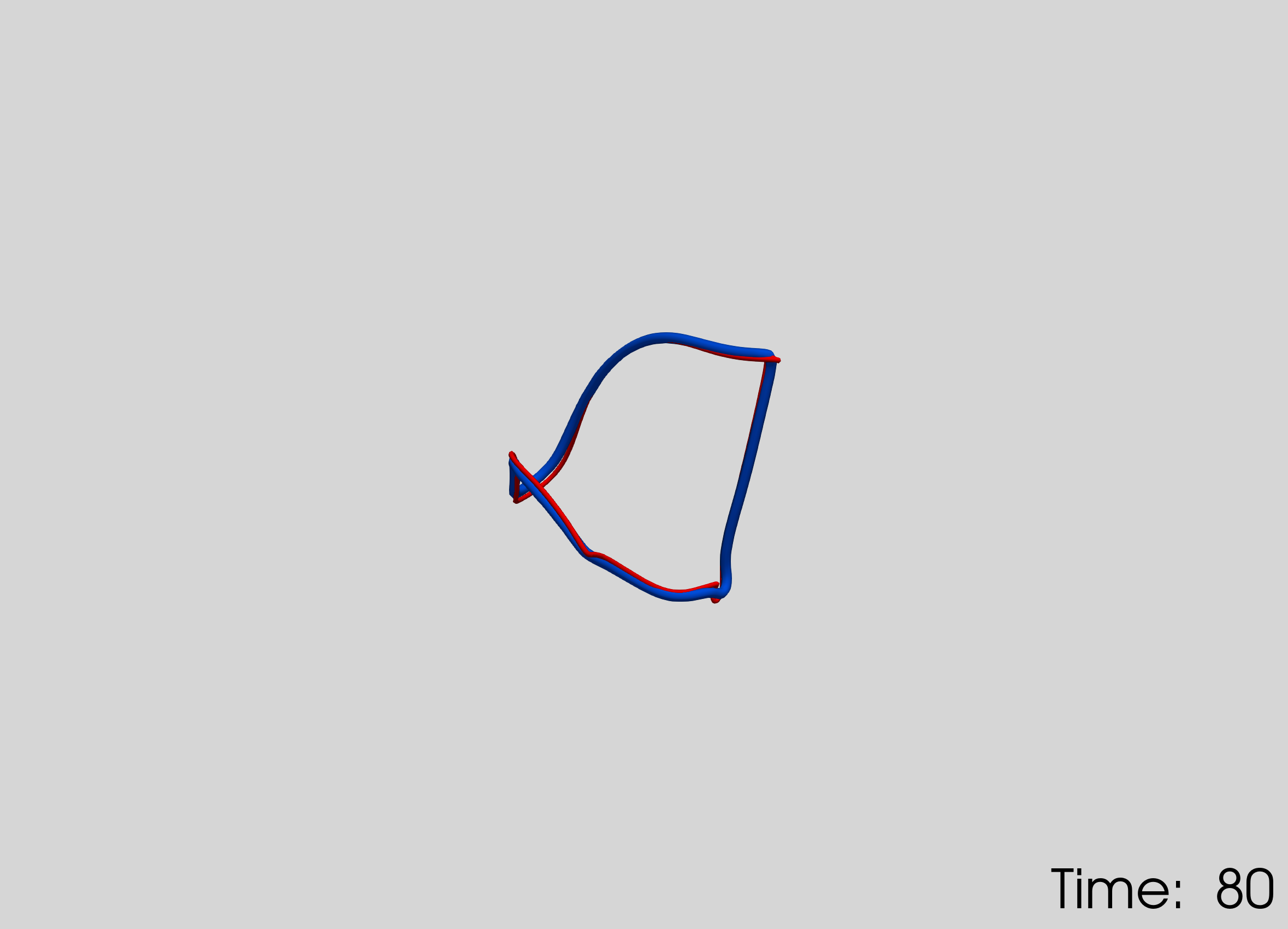}
\includegraphics[width=6.cm, height=4.2cm]{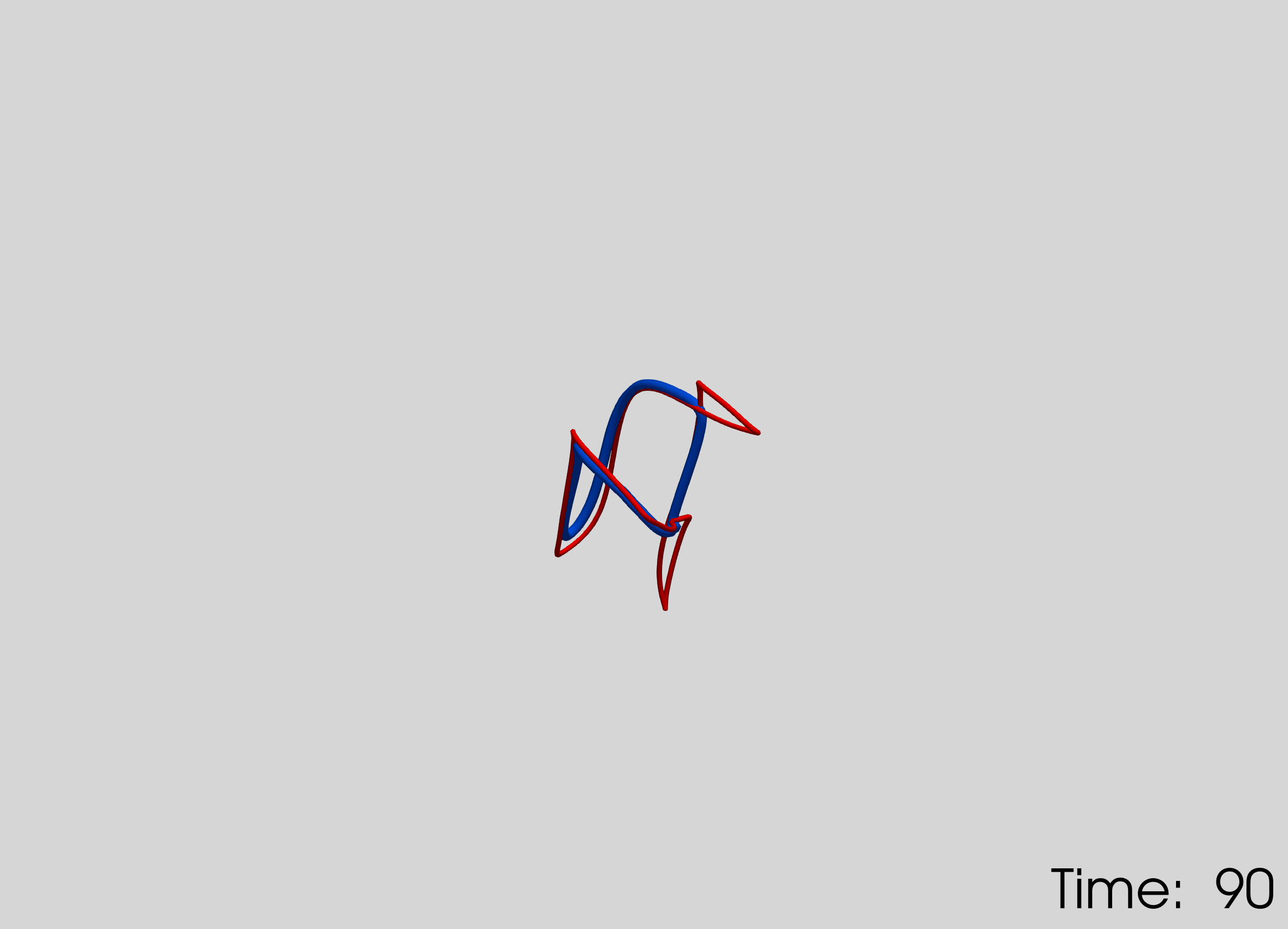}
\caption{Snapshots of the evolution of a primordial loop. The loop starts from
rest. We notice how the evolution of the field theory (in blue) is very close
to the NG (red) except towards the end of evolution where the loop
has shrunk by a large fraction and the NG predicts a complete overlap
of the extent of the string on itself.}
\label{fig:primordial-loops}
\end{center}
\end{figure}

We show in figure~\ref{fig:primordial-loops}
the comparison of the evolution of these primordial loops with their
predicted NG dynamics. The pictures demonstrate that the evolution is
pretty much identical in both cases all the way until moments before
the predicted overlap of the loop. There are no departures from the
NG dynamics due to high curvature regions for much of its
evolution. This is easy to understand since the loop is indeed much
smoother due to the diffusion period.  However, as the loop shrinks, we
start seeing some deviations from the NG behaviour, although not so dramatic
as in the non-primordial loops. As the loop comes close to its overlap, the 
difference between both field theory and NG becomes more apparent. This 
is also to be expected since the interaction of different regions of the string 
in this pathological self-intersection is of course not handled by the NG
dynamics. Nevertheless, the fact that up to this point both
descriptions agree with one another can be seen as a validation of
both the field theory and NG reconstruction codes.

\begin{figure}
\begin{center}
\includegraphics[width=12cm, height=8cm]{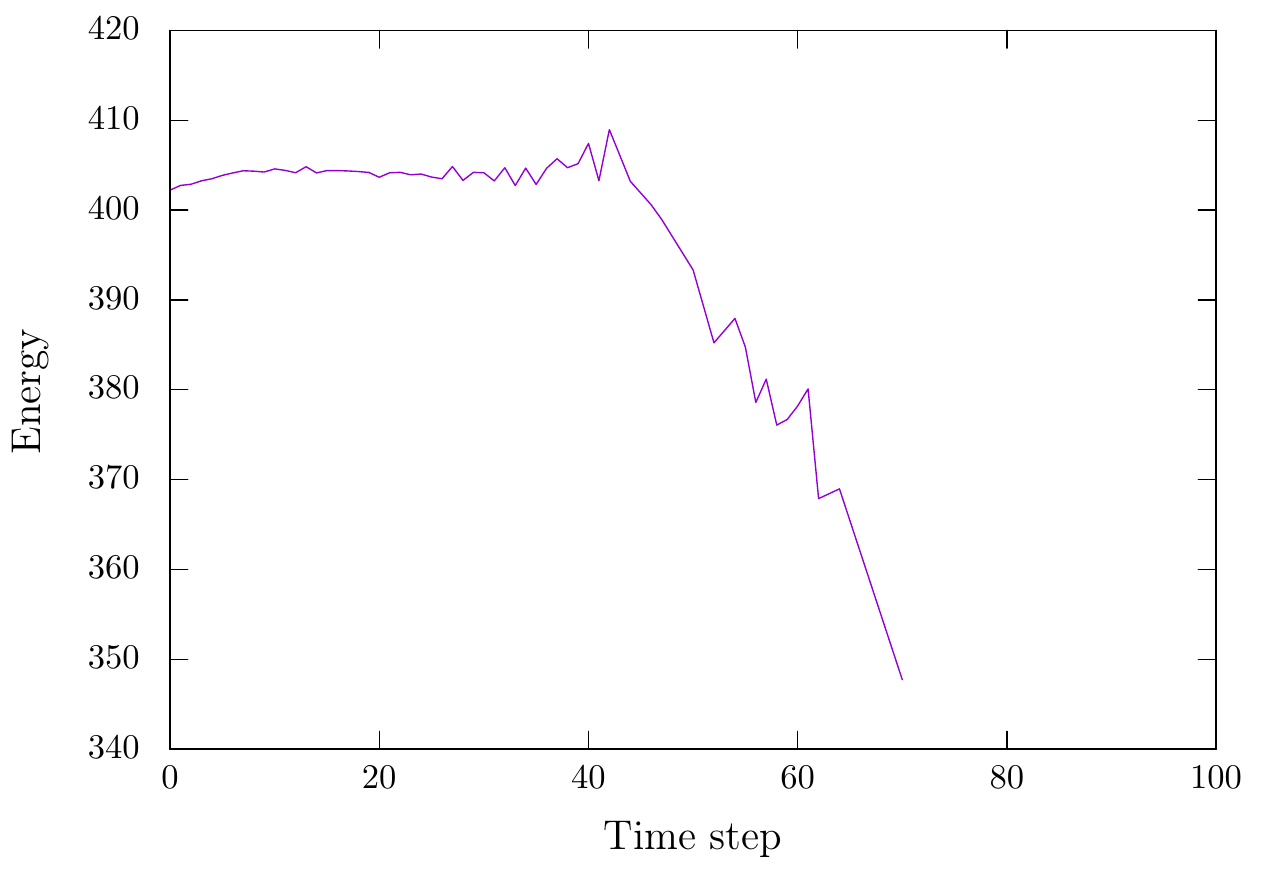}
\caption{Total amount of invariant energy (total amount of $\sigma$) for the
NG reconstruction of the field theory data of the primordial loop. }
\label{fig:NG-energy-reconstructed-primordial-loop}
\end{center}
\end{figure}

We also show in figure~\ref{fig:NG-energy-reconstructed-primordial-loop}
the energy of this loop using the NG reconstruction at each moment
in time. We notice that the energy is pretty much constant until $t\sim 40$,
which is also the moment where there is the first signal of deviation from 
NG dynamics in the loop's evolution (see figs. \ref{fig:primordial-loops}).
The deviation from the NG prediction happens in several places in the loop
and even though they are associated with high curvature regions, they are not
so obvious, at least not visually, as the ones presented earlier in the non-primordial
loop. This is somewhat similar to what happens in figure~\ref{fig:last}. The energy of the loop is decreasing, and 
yet the dynamics of the loop seems to follow quite closely that of NG.

\bibliographystyle{JHEP}
\bibliography{FTvNG}

\end{document}